\documentclass[aps,prd,twocolumn,superscriptaddress,amsmath,amssymb,nofootinbib]{revtex4}
\usepackage{graphicx} 
\usepackage{epstopdf} 
\usepackage{dcolumn} 
\usepackage{xcolor} 
\usepackage{amsmath,amssymb} 
\usepackage{hyperref} 
\usepackage[utf8]{inputenc} 
\usepackage[english]{babel} 
\usepackage[displaymath, mathlines]{lineno} 
\usepackage{blindtext} 
\usepackage{ifthen} 
\usepackage{orcidlink} 

\usepackage{breakurl}
\usepackage{array}
\usepackage{slashed}
\usepackage{xspace}
\usepackage{booktabs}
\usepackage{hhline}
\usepackage{multirow}
\usepackage{placeins}
\usepackage{physics}
\usepackage{listings}
\usepackage{gensymb}
\usepackage{todonotes}
\usepackage{siunitx}
\usepackage{subfigure}
\usepackage{float}

\graphicspath{{figures/}} 

\newboolean{articletitles}
\setboolean{articletitles}{true} 

\input{belle2-symbols}

\ifthenelse{\isundefined{\analysis}}{
    
}{

}

\begin{document}

\begin{flushright}
Belle II Preprint: 2026-001\\
KEK Preprint: 2025-38\\
\end{flushright}

\title{Study of $B^+ \to \mu^+ \nu_\mu$ decays at Belle and Belle~II}
  \author{M.~Abumusabh\,\orcidlink{0009-0004-1031-5425}} 
  \author{I.~Adachi\,\orcidlink{0000-0003-2287-0173}} 
  \author{K.~Adamczyk\,\orcidlink{0000-0001-6208-0876}} 
  \author{A.~Aggarwal\,\orcidlink{0000-0002-5623-3896}} 
  \author{L.~Aggarwal\,\orcidlink{0000-0002-0909-7537}} 
  \author{H.~Ahmed\,\orcidlink{0000-0003-3976-7498}} 
  \author{Y.~Ahn\,\orcidlink{0000-0001-6820-0576}} 
  \author{H.~Aihara\,\orcidlink{0000-0002-1907-5964}} 
  \author{N.~Akopov\,\orcidlink{0000-0002-4425-2096}} 
  \author{S.~Alghamdi\,\orcidlink{0000-0001-7609-112X}} 
% ...
\collaboration{The Belle and Belle II Collaborations}

\begin{abstract}
We report a measurement of the branching fraction for the leptonic decay $B^+\to\mu^+\nu_\mu$. This work presents the first $B^+\to\mu^+\nu_\mu$ result using Belle~II data, an updated Belle measurement that supersedes the previous result, and their combination, which yields the most precise search to date. The analysis is based on $1076\,\mathrm{fb}^{-1}$ of $e^+e^-$ collision data collected at a center-of-mass energy of $10.58\,\mathrm{GeV}$ with the Belle and Belle~II detectors at the KEKB and SuperKEKB colliders, respectively. We measure $\mathcal{B}(B^+\to\mu^+\nu_\mu)=(4.4\pm1.9\pm 1.0)\times10^{-7}$, where the first uncertainty is statistical and the second systematic. The observed significance relative to the background-only hypothesis is 2.4 standard deviations. We set a 90\% confidence level upper limit of $\mathcal{B}(B^+\to\mu^+\nu_\mu)<6.7\times10^{-7}$ using a frequentist approach and a 90\% credibility level upper limit of $\mathcal{B}(B^+\to\mu^+\nu_\mu)<7.2\times 10^{-7}$ using a Bayesian approach. These are the most stringent limits to date. The result is interpreted as an exclusion region in the parameter space of type~II and type~III two-Higgs-doublet models. We search for stable sterile neutrinos with masses $m_N\in[0,1.5]\,\mathrm{GeV}$. No signal is observed, and the resulting exclusion on the squared mixing parameter $|U_{\mu N}|^2$ provides improvement over previous limits. We report a measurement of the partial branching fraction of semileptonic $B\to X_u\ell\nu_\ell$ decays with $p_\mu^B>2.2\,\mathrm{GeV}$, obtaining $\Delta\mathcal{B}(B\to X_u\ell\nu_\ell)=(2.72\pm0.05\pm0.29)\times10^{-4}$. We present a model-dependent study of weak annihilation decays using the muon momentum spectrum. We observe a signal of 2.4 standard deviations above the background-only hypothesis in regions where the distribution resembles that of $B\to X_u\ell\nu_\ell$ decays.
\end{abstract}

\maketitle
\section{Introduction}

\begin{figure}[htbp]
  \centering
  \subfigure[]{\includegraphics[width=0.48\linewidth]{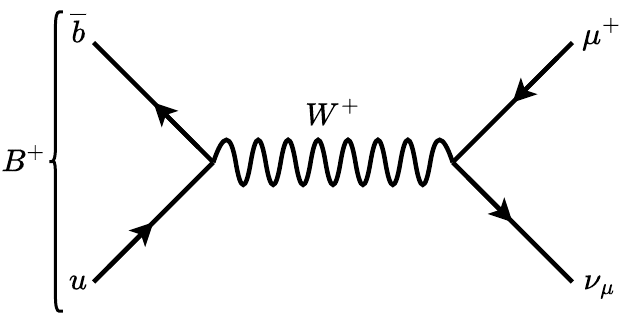}\label{fig:feynman_a}}\hfill
  \subfigure[]{\includegraphics[width=0.48\linewidth]{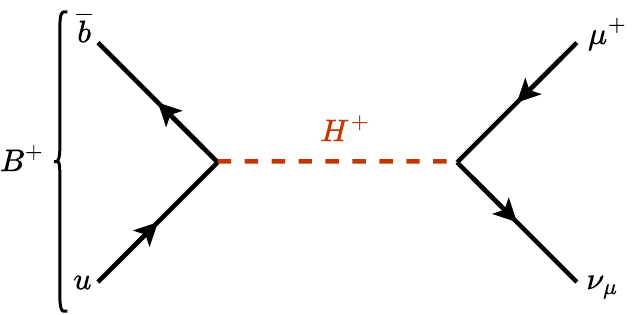}\label{fig:feynman_b}}
  \vskip\baselineskip
  \subfigure[]{\includegraphics[width=0.48\linewidth]{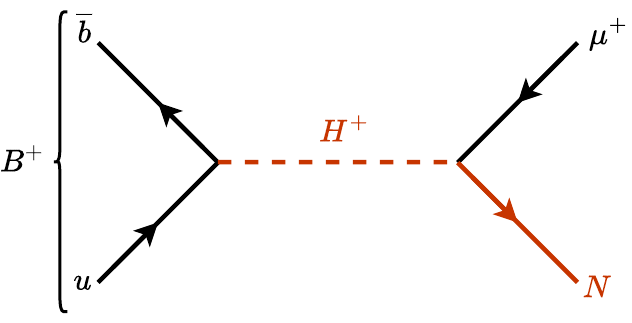}\label{fig:feynman_c}}\hfill
  \subfigure[]{\includegraphics[width=0.48\linewidth]{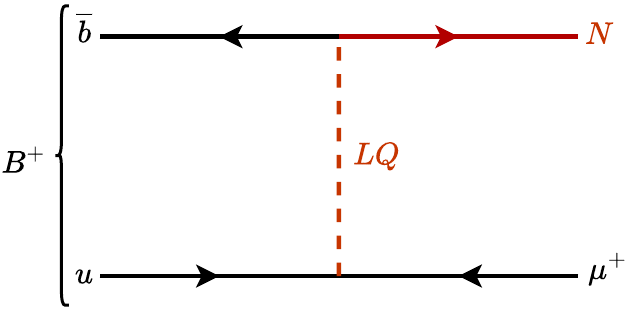}\label{fig:feynman_d}}
  \caption{Tree-level Feynman diagrams for the $B^+ \to \mu^+ \nu_\mu$ decay in the SM (a), and for potential beyond Standard Model contributions mediated  by a charged Higgs boson \(H^+\) ((b),(c)) or a leptoquark \(LQ\) (d). Note the final-state sterile neutrino \(N\) in cases (c) and (d).}
  \label{fig:Feynmans}
\end{figure}

Precision measurements of leptonic decays of \( B \) mesons, such as $B^+ \to \mu^+ \nu_\mu$, offer a unique way to test the validity of the Standard Model (SM) of particle physics. In the SM the $B^+ \to \mu^+ \nu_\mu$ decay proceeds via the annihilation of the constituent \( \bar b \) and \( u \) quarks into a virtual \( W^+ \) boson, which subsequently decays into an anti-lepton and the corresponding neutrino,\footnote{Charge conjugation and natural units $c = \hbar = 1$ are implied throughout.} as illustrated by the tree-level Feynman diagram in Fig.~\ref{fig:Feynmans} (a).

The SM branching fraction is given by 
\begin{equation}\label{eq:munu_bf}
    \mathcal{B}(B^+ \to \mu^+ \nu_\mu) = \frac{G_F^2 m_B m_\mu^2}{8 \pi} \left( 1 - \frac{m_\mu^2}{m_B^2} \right)^2 f_B^2 \left|V_{ub} \right|^2 \tau_{B^+} \, ,
\end{equation} 
where \( G_F \) is the Fermi constant, \( m_\mu \) is the muon mass, \( m_B = \left(5279.42 \pm 0.08 \right) \, \mathrm{MeV} \)~\cite{PDG_2024}, \( f_B = \left(190.0 \pm 1.3 \right) \, \mathrm{MeV} \)~\cite{FLAG_2024} and \( \tau_B = \left( 1.638 \pm 0.004  \right)\, \mathrm{ps} \)~\cite{HFLAV_2024} are the $B^+$ meson mass, decay constant and lifetime, and \(V_{ub}\) is the relevant Cabibbo--Kobayashi--Maskawa (CKM) matrix element. The decay is helicity suppressed and further suppressed by the small magnitude of \(\left|V_{ub} \right|\). In the SM, the branching fraction of $B^+ \to \mu^+ \nu_\mu$
is predicted to be
\begin{equation} \label{eq:SM_BF}
    \mathcal{B}(B^+ \to \mu^+ \nu_\mu) = (4.18 \pm 0.44) \times 10^{-7} \, ,
\end{equation}
using \( |V_{ub}| = (3.82 \pm 0.20) \times 10^{-3} \), obtained from the average of the world averages of inclusive and exclusive determinations~\cite{PDG_2024}.

In the presence of new interactions or particles beyond those in SM, the CKM and helicity suppression may be lifted, as illustrated by Fig.~\ref{fig:Feynmans} (b--d). The existence of a charged Higgs boson, as predicted in many supersymmetric extensions of the SM, could significantly modify the branching fraction~\cite{Higgs_Type_2,Higgs_Type_2_a,Higgs_Type_2_b}. Similar effects may also arise from leptoquarks~\cite{Hou:2019wiu}. Sterile neutrinos, as gauge singlets, could be produced in $B^+ \to \mu^+ N$ decays via a non-SM mediator, provided the sterile neutrino mass is smaller than $m_B - m_\mu$~\cite{Yue:2018hci}. 

Experimentally, the study of the $B^+ \to \mu^+ \nu_\mu$ decay is challenging due to its signature of a single muon and the presence of missing energy due to the neutrino. The most recent measurement of the leptonic $B^+ \to \mu^+ \nu_\mu$ decay by Belle~\cite{Prim_2020} reports a branching fraction of \( \mathcal{B}(B^+ \to \mu^+ \nu_\mu) = (5.3 \pm 2.0 \pm 0.9) \times 10^{-7} \), where the errors denote the statistical and systematic uncertainties, respectively, with a reported significance of 2.8 standard deviations above the background-only hypothesis. 

In this paper, we improve the precision of the measurement by combining Belle and Belle~II data sets, collected at the KEKB and SuperKEKB \(e^+e^-\) colliders~\cite{Kurokawa:2001nw,Abe:2013kxa,Akai:2018mbz} from collisions at and near the $\Upsilon(4S)$ resonance. The Belle and Belle~II samples correspond to integrated luminosities of \SI{711}{\per\femto\barn} and \SI{365}{\per\femto\barn}, respectively. The $\Upsilon(4S)$ decays almost exclusively to \(B\bar{B}\) pairs. The corresponding numbers of $\Upsilon(4S)$ events are \( (772 \pm 10)\times 10^6 \) and \( (387 \pm 6)\times 10^6 \), respectively, from which the numbers of \(B\) meson pairs are determined as \mbox{\(N_{\Upsilon(4S)} \times (f_{00} + f_{+-})\)}, where \(f_{00}\) and \(f_{+-}\) are the fractions of neutral and charged \(B\) pairs produced. Assuming the branching fraction in Eq.~\ref{eq:SM_BF}, about 500 \mbox{$B^+ \to \mu^+ \nu_\mu$} decays are expected in the combined data set. To maximize efficiency, we employ an inclusive tagging approach. While hadronic and semileptonic tagging methods are, in principle, applicable, they are not yet competitive given the low number of expected decays. For the Belle measurement, input parameters relevant to background modeling, such as inclusive and exclusive branching fractions and form factors, have been updated to improve the overall description of the data.

The signal $B^+ \to \mu^+ \nu_\mu$ decay is reconstructed by identifying a high-momentum muon. The remaining particles in the collision event, which include all reconstructed charged particles and photons, define the rest of the event (ROE) and are used to reconstruct the tag-side \(B\) meson. Exploiting the back-to-back kinematics of \(\Upsilon(4S)\to B\bar{B}\) decays in the \(e^+e^-\) center-of-mass (c.m.)\ frame, the four-momentum of the signal-side \(B\) is determined from the tag-side reconstruction. This allows the analysis of the muon in the approximate signal-\(B\) rest frame, where its momentum shows a peak around the expected monochromatic value of \(p_\mu^B = \SI{2.64}{\giga\eV}\) from the two-body decay.

The remainder of this paper is organized as follows: Section~\ref{ref:data_set} introduces the data sets, simulations, and applied corrections. Updates to the Belle samples from the previously published analysis are detailed in Section~\ref{ref:Belle_upd}. The overall analysis strategy, calibration of the ROE, and methods used to suppress the background from continuum processes are outlined in Section~\ref{strategy}. The various analysis steps are validated using the control $B^+ \to \bar D^0 \pi^+$ decay, as discussed in Section~\ref{re:CC_text}. The statistical methods used to determine the branching fraction of the $B^+ \to \mu^+ \nu_\mu$ decay are presented in Section~\ref{Fit_setup}, followed by a discussion of systematic uncertainties in Section~\ref{ref:sys_uncerts}. The results are presented in Section.~\ref{ref:results}. Additionally, the analysis allows for a precise measurement of the partial branching fraction of semileptonic $B \to X_u \mu^+ \nu_\mu$ decays at the kinematic end point ($p_\mu^B > \SI{2.2}{\giga \eV}$) and provides sensitivity to weak annihilation~(WA) processes. This is discussed in Sections~\ref{ref:ulnu_partial} and \ref{ref:wa_proc}, respectively. We conclude with Section~\ref{ref:conclusion}.

\section{Detector, simulated samples and corrections}\label{ref:data_set}

\subsection{Belle~II Detector}\label{ref:Belle_detector}

The Belle and Belle~II detectors are general-purpose spectrometers. A detailed description of the Belle detector can be found in Refs. \cite{Prim_2020,ABASHIAN2002117}. In the following, a brief overview of the Belle~II detector is provided, with further details available in Ref. \cite{B2physbook}.

The Belle~II detector consists of several subdetectors arranged cylindrically around the \(e^+e^-\) interaction point~\cite{Abe:2010gxa}. Charged-particle trajectories (tracks) are reconstructed using the vertex detector, comprising a two-layer silicon-pixel detector (PXD) and a four-layer double-sided silicon-strip detector (SVD)~\cite{Belle-IISVD:2022upf}, together with a 56-layer central drift chamber (CDC), which also provides $\text{d}E/\text{d}x$ measurements. For the analyzed data, only one sixth of the second PXD layer was installed. Particle identification is achieved with a time-of-propagation counter in the barrel region~\cite{Kotchetkov:2018qzw} and an aerogel ring-imaging Cherenkov detector in the forward endcap. Outside these, an electromagnetic calorimeter (ECL) based on CsI(Tl) crystals measures the energy and timing of electrons and photons. A superconducting solenoid surrounds these components, providing a 1.5~T magnetic field parallel to the electron beam, essential for momentum and charge measurements. The outermost subsystem, the instrumented flux return, detects \(K_L^0\) mesons, muons, and neutrons using resistive-plate chambers and plastic scintillator modules. The detector is described in a right-handed Cartesian coordinate system with the origin at the interaction point (IP). The \(z\)-axis is defined as the Belle~II solenoid axis, the \(x\)-axis points outward, and the \(y\)-axis upward.

\subsection{Belle~II Simulation}\label{ref:Belle2_upd}

Monte Carlo (MC) simulations are utilized to determine reconstruction efficiencies, to account for acceptance effects, to assess background contamination, and to provide training samples for multivariate classifiers. The simulation of \( B \) meson decays is performed using \texttt{EvtGen}~\cite{evtgen} interfaced to \texttt{PYTHIA8}~\cite{pythia} and electromagnetic final-state radiation is simulated using the \texttt{PHOTOS} package~\cite{Barberio:1990ms} for all charged final-state particles. Interactions of particles with the detector are simulated using \texttt{GEANT4}~\cite{GEANT4:2002zbu}. All recorded $e^+e^-$ collision data and simulated events are reconstructed and analyzed with the open-source \texttt{basf2} framework~\cite{Kuhr:2018lps,belle2_basf2_2021}. Templates in the form of binned histograms are constructed from these simulations to model the distributions of signal and background events. Unless stated otherwise, the template for $B^+ \to \mu^+ \nu_\mu$ decays is normalized to the expected number of events, assuming the branching fraction given by Eq. \ref{eq:SM_BF}. The backgrounds considered are \(B^+ \to \mu^+ \nu_\mu \gamma\) decays; semileptonic \(b \to u\) and \(b \to c\) transitions; other rare \(B\) decays; continuum processes \( e^+e^- \to q \bar{q} \) where \( q = \{u,d,s,c\} \); \( e^+e^- \to \ell^+ \ell^- (\gamma) \) processes with \( \ell = \{e,\mu,\tau\} \); as well as \( e^+e^- \to \gamma \gamma (\gamma) \); and four fermion processes \( e^+e^- \to e^+ e^- e^+ e^- \) and \( e^+e^- \to e^+ e^- \mu^+ \mu^- \). The simulation of the \( e^+e^- \to q \bar{q} \) processes is performed with \texttt{KKMC}~\cite{kkcm}, using \texttt{PYTHIA8} for quark fragmentation. The processes \( e^+e^- \to \mu^+ \mu^- (\gamma) \) and \( e^+e^- \to \tau^+ \tau^- (\gamma) \) are also generated with \texttt{KKMC}, with \(\tau\) decays handled by \texttt{TAUOLA}~\cite{Jadach:1990mz}. Quantum electrodynamics (QED) processes such as \( e^+e^- \to e^+ e^- (\gamma) \) and \( e^+e^- \to \gamma \gamma (\gamma) \) are simulated using \texttt{BABAYAGA.NLO}~\cite{BABAYAGA_1,BABAYAGA_2,BABAYAGA_3,BABAYAGA_4,BABAYAGA_5}, while four-fermion processes, \( e^+e^- \to e^+ e^- e^+ e^- \) and \( e^+e^- \to e^+ e^- \mu^+ \mu^- \), are modeled with \texttt{AAFH}~\cite{aafh_1,aafh_2,aafh_3}.

At next-to-leading order in QED, the $B^+ \to \mu^+ \nu_\mu$ decay can be accompanied by photon emission, either from the final-state lepton or from the initial-state quarks. The soft radiation from the final-state lepton is included in the signal simulation using the \texttt{PHOTOS} package~\cite{Barberio:1990ms}. If the photon is instead emitted from the initial quarks, the helicity suppression of the decay is lifted; however, the rate is reduced by the fine structure constant, \(\alpha_{\mathrm{em}} \simeq 1/137\). This contribution is denoted as \(B^+ \to \mu^+ \nu_\mu \gamma\) and is treated as a separate background process. The most precise measurement of this branching fraction is 
\begin{equation}
    \Delta \mathcal{B}(B^+ \to \mu^+ \nu_\mu \gamma) = (1.00 \pm 1.08) \times 10^{-6} \, ,
\end{equation}  
for photons with energy \(E_\gamma^B > \SI{1.0}{\giga \eV}\) in the \(B\) meson rest frame~\cite{munugammabf}. This result is extrapolated to the kinematic region \(E_\gamma^B > \SI{300}{\mega \eV}\), using the model of Ref.~\cite{Korchemsky:1999qb} and assuming negligible contributions below this threshold, yielding 
\begin{equation}\label{eq:munugamma_bf}
    \mathcal{B}(B^+ \to \mu^+ \nu_\mu \gamma) = (1.41 \pm 1.51) \times 10^{-6} \, .
\end{equation}  

Semileptonic \( b \to u \) transitions are simulated using several exclusive channels and non-resonant contributions. The exclusive decays \( B \to \pi \ell^+ \nu_\ell \), \( B \to \rho \ell^+ \nu_\ell \), and \( B^+ \to \omega \ell^+ \nu_\ell \) are modeled with the Bourrely--Caprini--Lellouch (BCL) parametrization~\cite{Bourrely:2008za} as implemented in the \texttt{eFFORT} package~\cite{effort}, using input parameters from Ref.~\cite{Prim:2019cws}. The \( B^+ \to \eta \ell^+ \nu_\ell \) and \( B^+ \to \eta' \ell^+ \nu_\ell \) channels are described with light-cone sum rules from Ref.~\cite{Duplancic:2015zna}. A hybrid approach~\cite{hybrid} is adopted to combine the resonant channels with the non-resonant \( B \to X_u \ell^+ \nu_\ell \) contribution while maintaining reliable modeling in the signal region. The non-resonant contribution is described by the De Fazio--Neubert (DFN) model~\cite{DFN}, which incorporates the triple-differential decay rate into its shape-function formalism to reproduce inclusive predictions. The model has two free parameters: the $b$-quark mass in the Kagan–Neubert scheme~\cite{Kagan:1998ym}, $m_b^{\mathrm{KN}} = 4.658^{+0.103}_{-0.200}$, and the nonperturbative parameter $a^{\mathrm{KN}} = 1.328^{+0.617}_{-0.704}$. Their values were determined in Ref.~\cite{Buchmuller:2005zv} from a fit to the spectral moments of $B \to X_c \ell \bar \nu_\ell$ and $B \to X_s \gamma$ decays. Hadronization of the parton-level \( B \to X_u \ell \bar{\nu}_\ell \) DFN simulation is performed with \texttt{PYTHIA8}~\cite{pythia}, producing final states containing two or more mesons. The hybrid model is constructed in three kinematic variables: the hadronic invariant mass \( M_X \), the squared momentum transfer \( q^2 = (p_\ell + p_\nu)^2 \), and the lepton energy in the \( B \) meson rest frame \( E_\ell^B \). The binning scheme is
\begin{align*}
    M_X &= [0, 1.4, 1.6, 1.8, 2, 2.5, 3, 3.5]~\text{GeV} \, , \\  
    E_\ell^B &= [0, 0.5, 1, 1.25, 1.5, 1.75, 2, 3]~\text{GeV} \, , \\ 
    q^2   &= [0, 2.5, 5, 7.5, 10, 12.5, 15, 25]~\text{GeV}^2 \, .  
\end{align*}  
Within this framework, the number of events in each bin \((i,j,k)\) in $M_X$, $E_\ell$, and $q^2$ is defined as
\[
    H_{ijk} = R_{ijk} + w_{ijk} \cdot I_{ijk} \, ,
\]
where \(R_{ijk}\) and \(I_{ijk}\) denote the resonant and non-resonant contributions, respectively. The weight factor \(w_{ijk}\) is chosen such that the partial rate in each hybrid bin \(H_{ijk}\) is equal to the partial rate in each fully inclusive bin \(I_{ijk}\), ensuring that the hybrid model is consistent with the inclusive prediction.  

\begin{figure}[htbp]
    \centering
    \includegraphics[width=\linewidth]{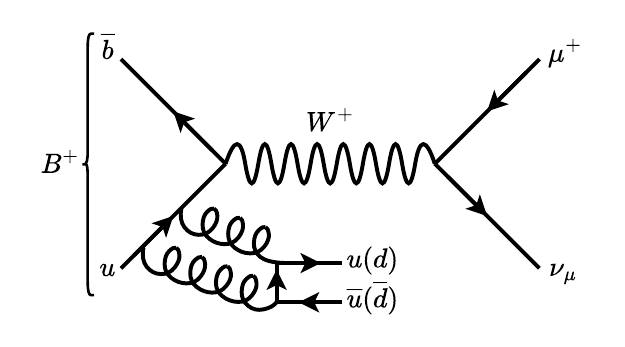}
    \caption{Leading-order Feynman diagram for the WA processes  \(B^+ \overset{\text{WA}}{\to} X_u \ell^+  \nu_\ell\). The hadronic system \(X_u\) is formed through initial-state gluon radiation, while the charged lepton and neutrino originate from the decay of the virtual \(W^+\) boson.}
    \label{fig:Feynman_WA}
\end{figure}

\begin{figure}[htbp]
    \centering
    \includegraphics[width=\linewidth]{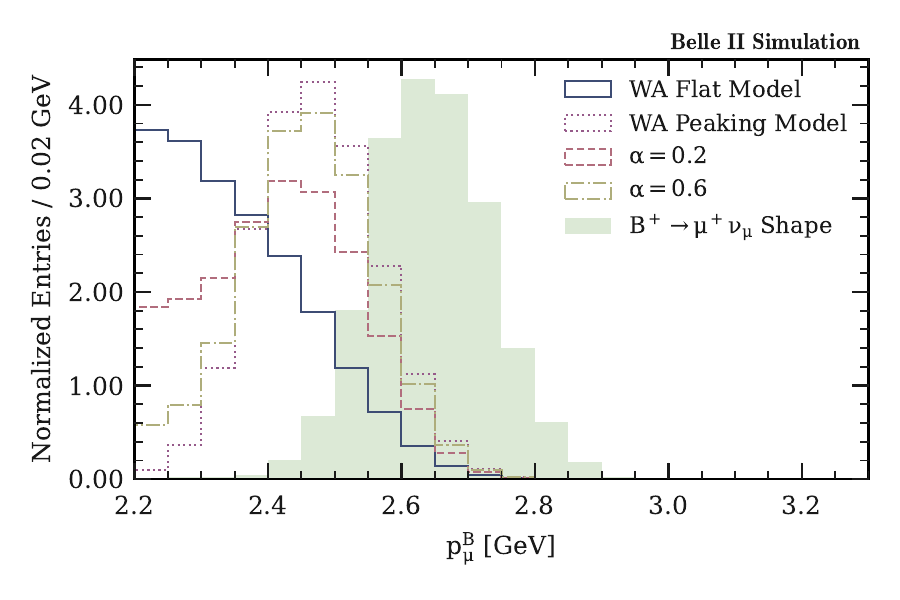}
    \caption{Distribution of the muon momentum $p_\mu^B$ in the \(B\) meson rest frame for the flat, peaking and intermediate \(B^+ \overset{\text{WA}}{\to} X_u \ell^+  \nu_\ell\) models, with the transition between shapes controlled by the parameter \(\alpha\), compared to that of $B^+ \to \mu^+ \nu_\mu$ decays.}
    \label{fig:WA_Shapes}
\end{figure}

In WA processes, initial-state gluon radiation produces a bound state \(X_u\), while the \(b\) and \(u\) quarks annihilate into a virtual \(W\) boson that subsequently decays into a lepton and a neutrino.  
The tree-level Feynman diagram for this process is shown in Fig.~\ref{fig:Feynman_WA}.  Such decays can closely resemble $B^+ \to \mu^+ \nu_\mu$ events, and both their kinematic distributions and their relative contribution to the semileptonic \(b \to u\) rate are poorly understood. Experimentally the relative contribution with respect to the full $B \to X_u \ell \bar \nu_\ell$ rate is constrained to be below 7.4--13\% at 90\% CL~\cite{Belle:2021eni,BaBar:2011xxm,CLEO:2006xbj}, whereas theoretical limits from $D$ and $D_s$ decays estimate a relative contribution of the order of 1--3\%~\cite{Gambino:2010jz,Ligeti:2010vd,Voloshin:2001xi,Bigi:1993bh}. We model the WA decays as a sequence of two-body decays, \(B^+ \to X_u W\) followed by \(W \to \ell^+ \nu_\ell\). The muon momentum spectrum \(p_\mu^B\) in the \( B \) meson rest frame is determined by the masses and widths of the hadronic system \( X_u \) and the virtual \( W \) boson. An admixture of two scenarios is considered: a flat distribution in \( p_\mu^B \), which arises from a broad spectrum of events at low \( q^2 \), and a peaking distribution, originating from a narrow resonance-like structure at high \( q^2 \). We implement a continuous interpolation between the two models, controlled by a parameter \(\alpha\) that scales their relative contributions. Figure~\ref{fig:WA_Shapes} shows the resulting \(p_\mu^B\) spectra of this model for flat ($\alpha = 0$), peaking ($\alpha = 1$), and intermediate cases. The distribution from $B^+ \to \mu^+ \nu_\mu$ decays is also shown for comparison. To convert the WA yields into a branching fraction, the overall efficiency is expressed as 
\begin{displaymath}
\epsilon = \epsilon_{\text{flat}} \cdot \left(1-\alpha \right) + \epsilon_{\text{peak}} \cdot \alpha,
\end{displaymath}
where $\epsilon_{\text{flat}}$ and $\epsilon_{\text{peak}}$ denote the efficiencies of the flat and peaking components, respectively.

The \( b \to c \) transitions are dominated by the exclusive decays 
\( B \to D \ell^+ \nu_\ell \) and \( B \to D^* \ell^+ \nu_\ell \), which are modeled with the Boyd--Grinstein--Lebed (BGL) parametrization~\cite{Boyd:1997kz, Grinstein:2017nlq, Bigi:2017njr}, using input parameters provided by the Heavy Flavor Averaging Group (HFLAV)~\cite{HFLAV_2024}. Decays to orbitally excited states, 
\( B \to D^{**} \ell^+ \nu_\ell \) with 
\( D^{**} = \{D_0^{*}, D_1^{\prime}, D_1, D_2^{*}\} \), are described using heavy-quark--symmetry--based Bernlochner--Ligeti--Robinson (BLR) form factors~\cite{Bernlochner:2016bci,Bernlochner:2017jxt}. The masses and widths of the \( D^{**} \) states are taken from Ref.~\cite{PDG_2024}, while the branching fractions are fixed to the HFLAV values~\cite{HFLAV_2024}, ensuring consistency with isospin-conjugated and other hadronic \( B \) decays, following the prescription of Ref.~\cite{Bernlochner:2016bci}. 

To account for the difference between the inclusive \( B \to X_c \ell^+ \nu_\ell \) rate and the sum of measured exclusive channels, we include ``gap processes'' modeled as intermediate broad resonances within the BLR framework. These serve only to ensure completeness and have negligible impact in the signal region. Finally, we also account for hadronic backgrounds from \( b \to c \) transitions in this contribution.

In addition to the dominant \( b \to c \) and \( b \to u \) transitions, we include rare hadronic processes, which are mainly given by \( B^+ \to K_L^0 \pi^+ \) and \( B^0 \to K_0^{*+} \pi^- \) decays.

The branching fractions for all decays are summarized in Table~\ref{tab:X_BF}.

\begin{table}[htbp]
    \caption{Branching fractions for various decays used in the analysis~\cite{PDG_2024,HFLAV_2024}. The WA branching fraction is taken as the average of the model-independent limits reported in Refs.~\cite{Belle:2021eni,BaBar:2011xxm}, corresponding to \((1 \pm 7)\%\) of the total \(B \to X_u \ell^+ \nu_\ell\) rate.}
\begin{tabular}{l | c|c}
\hline \hline
                        Decay Process     & $B^0$                                               & $B^+$                                                     \\ \hline    
\quad $B \to D \ell^+  \nu_\ell$            & $ (2.12 \pm 0.06) \times 10^{-2} $  & $ (2.21 \pm 0.06) \times 10^{-2} $ \\ 
\quad $B \to D^* \ell^+  \nu_\ell$          & $ (4.90 \pm 0.12) \times 10^{-2} $  & $ (5.53 \pm 0.22) \times 10^{-2} $  \\ 
\quad $B^+ \to \bar{D}^0 \pi^+$           &                                       & $ (4.66 \pm 0.14) \times 10^{-3} $  \\ 
\quad $B^0 \to D^- \pi^+$           & $ (2.48 \pm 0.11) \times 10^{-3} $  &                            \\ 
\hline                     
\quad $B^+ \to K^0 \pi^+$          &   & $ (2.39 \pm 0.06) \times 10^{-5} $ \\ 
\quad $B^0 \to K_0^{*+} \pi^-$            & $ (3.3 \pm 0.7) \times 10^{-5} $  &  \\ 
\hline                     
\quad $B^+ \to \mu^+ \nu_\mu \gamma$            &  & $ (1.41 \pm 1.51) \times 10^{-6} $ \\ 
\hline                    
\quad $B \to \pi \ell^+  \nu_\ell$    & $ (1.50 \pm 0.06) \times 10^{-4} $   & $ (7.80 \pm 0.27) \times 10^{-5} $ \\ 
\quad $B \to \rho \ell^+  \nu_\ell$   & $ (2.94 \pm 0.21) \times 10^{-4} $  & $ (1.58 \pm 0.11) \times 10^{-4} $ \\ 
\quad $B^+ \to \omega \ell^+  \nu_\ell$ &                                    & $ (1.19 \pm 0.09) \times 10^{-4} $ \\ 
\quad $B^+ \to \eta \ell^+  \nu_\ell$   &                                    & $ (3.5 \pm 0.4) \times 10^{-5} $ \\ 
\quad $B^+ \to \eta' \ell^+  \nu_\ell$   &                                    & $ (2.7 \pm 0.7) \times 10^{-5} $ \\ 
\quad $B \to X_u \ell^+  \nu_\ell$   &     $ (1.85 \pm 0.20) \times 10^{-3} $                       & $ (1.99 \pm 0.22) \times 10^{-3} $ \\ 
\hline                   
\quad $B^+ \overset{\text{WA}}{\to} X_u \ell^+  \nu_\ell$   &                          & $ (1.99 \pm 13.9) \times 10^{-5} $ \\ 
\hline               
\quad $B^+ \to \mu^+ \nu_\mu$            &  & $ (4.18 \pm 0.44) \times 10^{-7} $ \\ 
\hline \hline

\end{tabular}
\label{tab:X_BF}
\end{table}

\subsection{Updates to the Belle Simulation}\label{ref:Belle_upd}

Since the publication of the Belle result~\cite{Prim_2020}, several branching fractions and external parameters have been measured with improved precision~\cite{PDG_2024, HFLAV_2024}. The simulated Belle samples are therefore reweighted according to the updated inputs and models, which have been aligned with the Belle~II analysis to ensure a consistent treatment of common systematic uncertainties. The branching fraction of the \(B^+ \to \mu^+ \nu_\mu \gamma\) decay has been updated to the value quoted in Eq.~\ref{eq:munugamma_bf}. For semileptonic \( b \to c \) transitions, in particular \( B \to D \ell^+ \nu_\ell \) and \( B \to D^* \ell^+ \nu_\ell \), the form factor parameters and corresponding branching fractions have been updated. Similarly, the branching fractions and form factor inputs for the dominant semileptonic \( b \to u \) transitions have been updated. In addition, the modeling of inclusive \( B \to X_u \ell^+ \nu_\ell \) decays has been updated to use consistent input parameters. 

\section{Event Selection}\label{strategy}

\subsection{Signal and Tag-side Reconstruction}

The selection criteria for the Belle data remain unchanged and are summarized in Ref.~\cite{Prim_2020}. The Belle~II analysis conceptually follows the same strategy and is outlined below.

The signal-\(B\) meson is reconstructed by selecting a muon with a momentum in the c.m.\ frame of the \( e^+e^- \) collision within the range of \( 2.1 \, \text{ GeV} < p_\mu^* < 3.4 \, \text{ GeV} \). Here, and throughout the paper, quantities in the \( e^+e^- \) center of mass frame are indicated by an asterisk. Muons are identified using the likelihood ratio \( \mathcal{P}_{\mu} = \mathcal{L}_{\mu}/(\mathcal{L}_{e}+\mathcal{L}_{\mu}+\mathcal{L}_{\pi}+\mathcal{L}_{K}+\mathcal{L}_{p}+\mathcal{L}_{d})\), where the likelihood \(\mathcal{L}_i\) for each charged-particle hypothesis ($i = e, \mu, \pi, K, p, d$) combines particle identification information from all detectors except the silicon trackers. By construction, \(\mathcal{P}_{\mu}\) ranges from zero to one. A requirement of \( \mathcal{P}_{\mu} > 0.99 \) is applied to ensure high-purity muon identification. Within the specified momentum range, this requirement selects 85\% of muons and removes 98.5\% of particles that are not muons. Additionally, the candidate must originate from the IP, enforced through a constraint on the point of closest approach along the beam axis \(|\mathrm{d}z| < 2 \text{ cm}\), and the transverse distance \(\mathrm{d}r < 0.5 \, \text{ cm}\). Only events with more than one charged track in the ROE are considered. Each charged particle in the ROE is assigned its most-likely mass hypothesis by selecting the particle species that maximizes the respective particle-identification likelihood ratio. The resulting multiplicities of each particle type are then constrained to match the expected distributions obtained from simulated \( \Upsilon(4S) \) decays, ensuring that the overall composition of particle species in the ROE reflects the physical expectations. Charged particles must satisfy the constraints \(|\mathrm{d}z| < 3 \, \text{ cm}\) and \(\mathrm{d}r < 0.5 \text{ cm}\), the transverse momentum requirement \(p_T > 100 \, \text{ MeV}\), and their energy $E$ must not exceed \(5.5 \text{ GeV}\). Furthermore, their trajectories must fall within the acceptance of the CDC, \(17\degree < \theta < 150\degree\), where \(\theta\) is the polar angle. Photons, reconstructed as neutral clusters in the ECL (energy deposits in the calorimeter not associated with any charged particle track), are considered if they meet the energy requirement \(0.075\, \text{ GeV}< E < 5.5 \, \text{ GeV}\), and are within the CDC acceptance. To ensure reliable reconstruction, only clusters from multiple crystals are retained. Additionally, clusters are required to satisfy the timing constraints 
\(|t_{\text{cluster}}| < 200\,\text{ns}\) and \(\left| t_{\text{cluster}}/\sigma_{t_{\text{cluster}}} \right| < 2.0\), where \(t_{\text{cluster}}\) denotes the difference between the photon timing 
and the event time, and \(\sigma_{t_{\text{cluster}}}\) is the estimated timing resolution, defined to contain approximately 99\% of true photons. To correct for discrepancies in photon reconstruction efficiency between data and MC, photons are removed from the reconstruction with a probability of \(1-w\), where \(w\) is the ratio of the efficiency in data to that in MC as determined with control sample studies~\cite{photon_eff,tau_nu_BF}.

The three-momentum of the tag-side $B$ meson in the laboratory frame is reconstructed as 
\begin{align}
    \vec p_{\text{tag}}^{\text{ lab}} 
    = \sum_{i \in \text{charged}} \vec p_i^{\text{ lab}} 
    + \sum_{j \in \gamma} \vec p_j^{\text{ lab}},
\end{align}  
where $ \vec p_i^{\text{ lab}}$ and $ \vec p_j^{\text{ lab}}$ are the three-momenta of the charged particles and photons in the ROE, respectively. The tag-side energy is then obtained from 
\begin{align}
    E_{\text{tag}}^{\text{lab}} 
    = \sqrt{ \bigl| \vec{p}_{\text{tag}}^{\text{ lab}} \bigr|^2 + m_{\text{tag}}^2 } \, ,
\end{align}
using the reconstructed mass of the tag-side \(B\) meson \(m_{\text{tag}}\). The normalized beam-constrained mass \(\hat{m}_{\mathrm{bc}}^{\text{tag}}\) and energy difference \(\Delta \hat{E}\) are defined as
\begin{align}
 \hat{m}_{\mathrm{bc}}^{\text{tag}} &= \frac{\sqrt{s/4 - \abs{\Vec{p}_{\text{tag}}^{\,*}}^2}}{\sqrt{s}/2} \, , \\
 \Delta \hat{E} &= \frac{E_{\text{tag}}^{*} - \sqrt{s}/2}{\sqrt{s}/2} \, ,  
\end{align}
where \( \sqrt{s} \) is the collision energy. Events must satisfy  
\begin{equation}
    \hat{m}_{\mathrm{bc}}^{\text{tag}} > 0.964, \quad -0.5 < \Delta \hat{E} < 0.1 \, ,
\end{equation}  
translating to an on-resonance requirement of \( m_{\mathrm{bc}}^{\text{tag}} = \sqrt{s/4 - \abs{\Vec{p}_{\text{tag}}^{\,*}}^2}\) greater than \( \SI{5.10}{\giga\electronvolt} \) and \(\Delta E = E_{\text{tag}}^{*} - \sqrt{s}/2 \) between \(\SI{-2.64}{\giga\electronvolt}\) and \(\SI{0.53}{\giga\electronvolt}\) in the \(\Upsilon(4S)\) rest frame. These requirements retain 62\% of reconstructed signal events while effectively suppressing 99.99\% of continuum processes.

\subsection{Tag-side Momentum Calibration}

\begin{figure}[htbp]
    \centering
    \includegraphics[width=\linewidth]{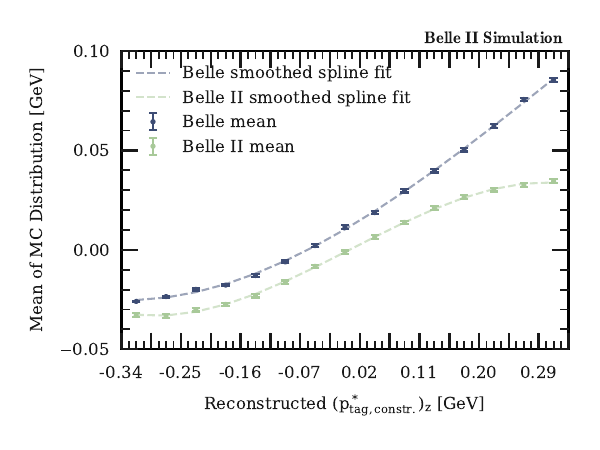}
    \caption{Relation between the mean generated and reconstructed values of the momentum component \( p_z \) in each bin. The difference between the Belle and Belle~II distributions stems from the different mass constraint enforced on the tag-side \(B\) meson.}
    \label{fig:Combined_Calibration_Plot}
\end{figure}

Exploiting the two-body kinematics of the \(\Upsilon(4S) \to B\bar{B}\) decay, the tag-side four-momentum is reconstructed as  
\begin{align}
    (p_{\text{tag,constr.}}^*)_i 
        &= (p_{\text{tag}}^*)_i \cdot 
           \frac{|\vec{p}_B^{\,*}|}{|\vec{p}_{\text{tag}}^{\,*}|}, \\
    E_{\text{tag,constr.}}^* 
        &= \sqrt{ |\vec{p}_B^{\,*}|^2 + m_B^2 } \, , \label{eq:Etag_B_exp}
\end{align}
where \(|\vec{p}_B^{\,*}|\) denotes the momentum of a charged \(B\) meson in the c.m.\ frame, with an expected value of 
\(\SI{332}{\mega\electronvolt}\), and with $i\in\{x,y,z\}$. The incomplete coverage of the \(4\pi\) solid angle and the presence of unreconstructed particles such as neutrinos and \(K^0_L\) mesons, limit the ROE reconstruction, which in turn affect the resolution of the muon momentum in the rest frame of the signal-\(B\) meson. To correct for these detector-level effects, a calibration function \(f\) is derived from simulated signal events. This function maps the reconstructed \(p_z\) component to the mean of the corresponding simulated distribution, as illustrated in Fig.~\ref{fig:Combined_Calibration_Plot}.

\begin{figure}[htbp]
    \centering
    \includegraphics[width=\linewidth]{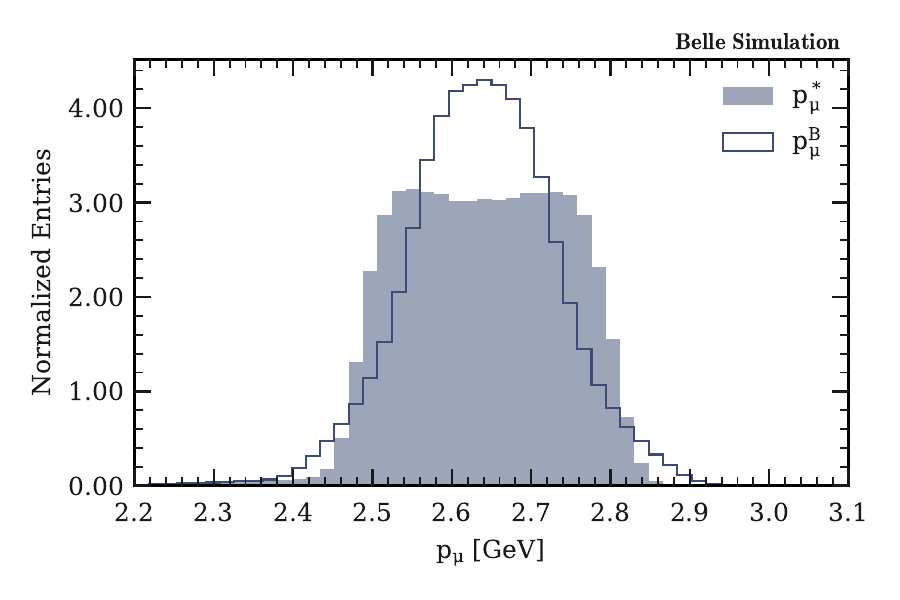}
    \includegraphics[width=\linewidth]{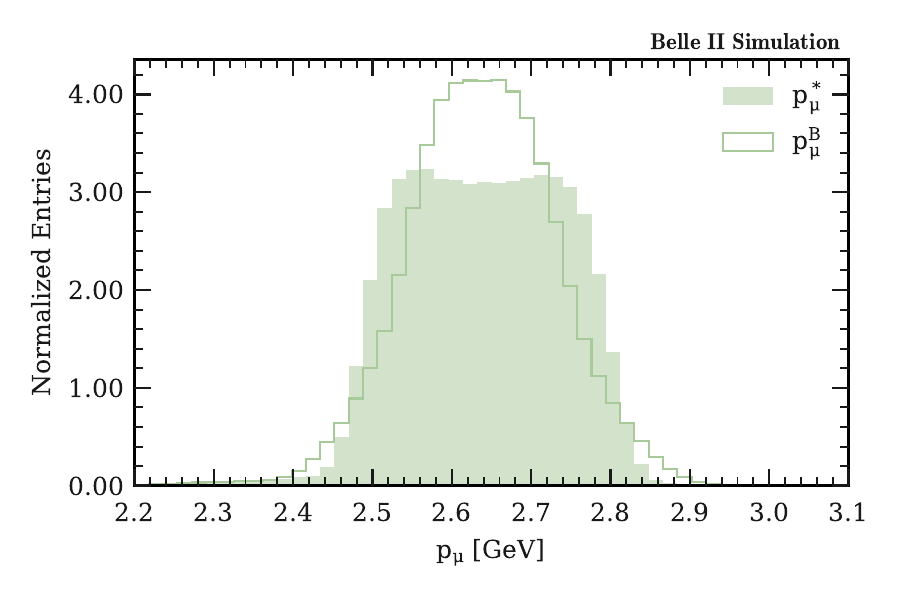}
    \caption{Muon momenta \(p_\mu^B\) and \(p_\mu^*\) in the estimated signal-\(B\) meson rest frame and in the c.m.\ frame, respectively, from the Belle (top) and Belle II (bottom) data.}
    \label{fig:Combined_pmuB}
\end{figure}

The corrected $p_z$ component is then used to determine the transverse momentum. An overall scaling factor \( \zeta \) is applied to minimize the difference between the corrected and generated three-momentum \(D = | \zeta \vec{p}^{\,*}_{\text{corr.}} - \vec{p}^{\,*}_{\text{gen.}} |\), resulting in:
\begin{align}
    (p_{\text{tag,\text{opt.}}}^{*})_z &= \zeta \cdot f \left[ (p_{\text{tag,constr.}}^*)_z \right], \\
    (p_{\text{tag,\text{opt.}}}^{*})_T &= \zeta \cdot \sqrt{\left( (p_{\text{tag,constr.}}^*) \right)^2 - \left( f \left[ (p_{\text{tag,constr.}}^*)_z \right] \right)^2} \, ,
\end{align}
with $(p_{\text{tag,\text{opt.}}}^{*})_T$ denoting the transverse-momentum component. The optimal scaling factor is found to be \( \zeta = 0.58 \) for the Belle data and \( \zeta = 0.52 \) for the Belle~II data. The optimized momentum components are combined with the energy component given by  Eq.~\ref{eq:Etag_B_exp}, allowing for the reconstruction of the signal-\( B \) meson kinematics:
\begin{equation}
     p^*_{\text{sig}} = \left( E_{\text{tag,constr.}}^*  , - \vec p_{\text{tag},\text{opt.}}^{\,*}  \right) \, .
\end{equation}
The signal muon momentum is boosted into the rest frame of the signal-\(B\) meson, leading to a resolution improvement of about 7\% and 6.4\% for Belle and Belle II, respectively, relative to the muon momentum in c.m.\ frame. The corresponding distributions are shown in Fig.~\ref{fig:Combined_pmuB}.

\subsection{Background Suppression}

Background contributions from continuum processes are suppressed using a multivariate classifier based on a boosted decision tree (BDT) trained with the \texttt{FastBDT} algorithm~\cite{fastBDT}, using event-shape observables as input variables (see e.g. Ref.~\cite{B2physbook}). These include the cosine of the angle between the signal muon momentum and thrust axis of the ROE, as well as harmonic moments of order zero to five with respect to the thrust axis of the event. The thrust axis is constructed from all tracks and neutral clusters in the object and is defined as the direction that maximizes the sum of the absolute values of the longitudinal momenta of the particles in the object. Additional input variables are the cosine of the polar angle of the missing momentum, \(\hat{m}_{\mathrm{bc}}^{\text{tag}}\), \(\Delta \hat{E}\), nine CLEO cones~\cite{Cleo_Cones}, and sixteen modified Fox--Wolfram moments~\cite{Belle:2003fgr}, which are constructed either exclusively from the ROE, denoted as type \textit{oo}, or from a combination of the ROE and the signal-\(B\) meson, denoted as type \textit{so}. The training process proceeds in three stages. 

First, a BDT is trained on the full set of 47 continuum-suppression variables using the off-resonance data and the signal MC. From this training, only the most important variables are retained, with the off-resonance data providing an unbiased representation of the continuum background. The off-resonance samples correspond to \SI{79}{\per\femto\barn} and \SI{42}{\per\femto\barn} of collision data collected \SI{60}{\mega\eV} below the \(\Upsilon(4S)\) resonance peak by the Belle and Belle~II detectors, respectively. 

In the second stage, the selected suppression variables, together with \(p_\mu^B\), are used to train a BDT with the off-resonance data against the continuum MC to correct for data–simulation differences. This BDT accounts for mismodelling in the continuum sample generated at the \(\Upsilon(4S)\) c.m.\ energy relative to the below-resonance data, taking into account the different collision energies. Its output, \(C_{\text{RW}}\), is converted into an event weight~\cite{Martschei_2012} 
\[
  w = \frac{C_{\text{RW}}}{1 - C_{\text{RW}}},
\]
which corrects both the shape and normalization of the simulated continuum events. Figure~\ref{fig:appendix_A_1} in Appendix~\ref{sec:appendix_A_1} shows the effect of this correction on a set of variables.

Before the final step, all selected variables are checked for agreement between off-resonance data and simulation. Variables that are highly discriminating but poorly modeled are removed, and, if necessary, strongly correlated alternatives are included to maintain the separation power. Further, any variables that show a correlation of more than 10\% to \(p_\mu^B\) are removed. This ensures that the final BDT uses only variables that both maximize signal–background discrimination and are reliably modeled, while avoiding any sculpting of \(p_\mu^B\).

\begin{figure}[htbp]
    \centering
    \includegraphics[width=\linewidth]{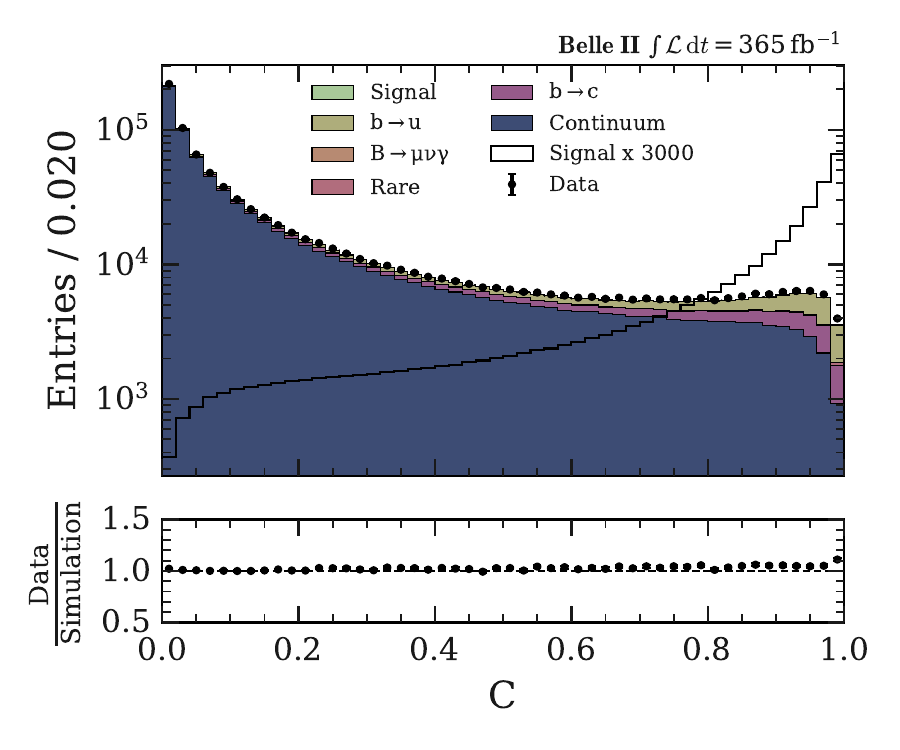}
    \includegraphics[width=\linewidth]{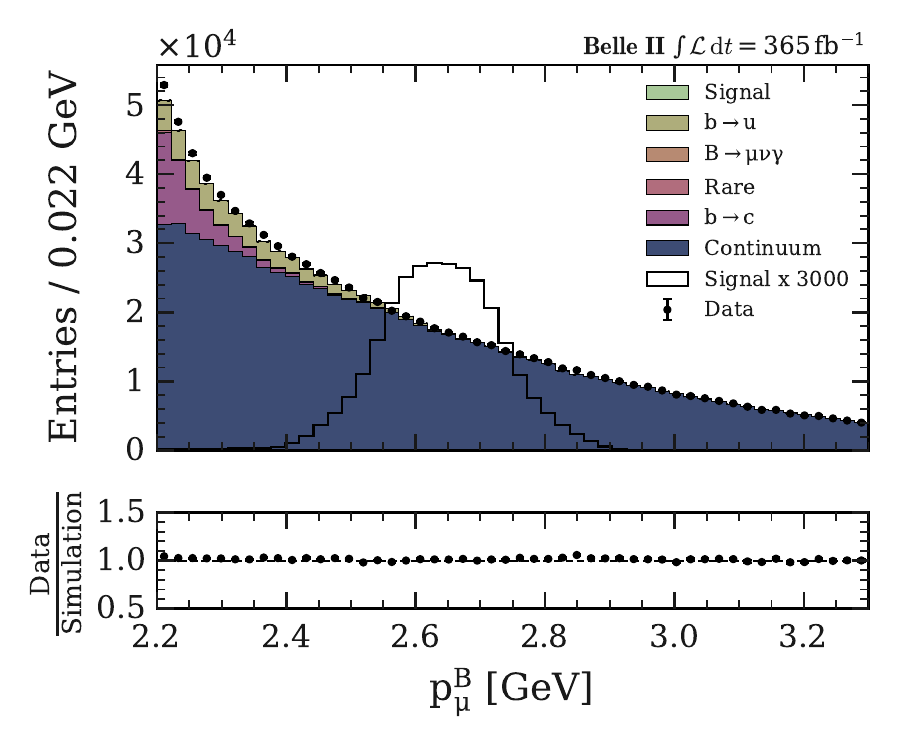}
    \caption{Distribution of the continuum suppression classifier output \(C\) (top) and the muon momentum $p_\mu^B$ in the signal-\(B\) meson rest frame (bottom) from data and simulation. The colored stacked histograms represent the different components: the signal process $B^+ \to \mu^+ \nu_\mu$ in light green (not visible in this plot), with the same distribution scaled by a factor of 3000 in black, the \(b \to u\) transitions in tan, the \(B^+ \to \mu^+ \nu_\mu \gamma\) processes in golden brown, the rare $B$ processes in rose, the \(b \to c\) transitions in medium purple, and the corrected continuum events in dark blue.}
    \label{fig:BDT_Output_B2}
\end{figure}

In the last stage, sixteen well-modeled and discriminating variables are selected to train the continuum-suppression BDT. The used variables are listed in Appendix~\ref{sec:appendix_A_0}. The resulting classifier output score \(C\) and \(p_\mu^B\) for on-resonance data and simulation are shown in Fig.~\ref{fig:BDT_Output_B2}.

The $B^+ \to \mu^+ \nu_\mu$ branching fraction is extracted from a binned maximum-likelihood fit to \(p_\mu^B\). The fit is performed in predefined bins of the classifier output, chosen to maximize the sensitivity by separating the signal-enriched from the background-enriched categories.

The Belle analysis defined four mutually exclusive categories (labeled as Category~I–IV), consisting of two signal-enriched and two background-enriched regions used to constrain the \(b \to u\) transitions and the continuum processes. These categories are retained in the combined Belle and Belle~II analysis. 

For Belle~II, a similar strategy is implemented to define categories. A multi-dimensional grid search on the classifier output \(C\) is used to define one signal-enriched region and three background-enriched regions. Unlike the Belle analysis, no further separation based on \(\cos\theta_{B\mu}\), where \(\theta_{B\mu}\) is an angle-like variable defined by the calibrated signal-\(B\) meson momentum in the c.m.\ frame and the muon momentum in the \(B\) rest frame, is implemented due to the lower integrated luminosity. The selected categories for Belle~II are defined as: 
\begin{align}
\text{Category V}   \,&:\, 1.000 > C > 0.994, \nonumber \\
\text{Category VI}  \,&:\, 0.994 > C > 0.986, \nonumber \\
\text{Category VII} \,&:\, 0.986 > C > 0.968, \nonumber \\
\text{Category VIII}\,&:\, 0.968 > C > 0.900. \nonumber
\end{align}
Category~V corresponds to the signal-enriched region, while the remaining categories are used to constrain background contributions. Requiring \(C > 0.900\) yields selection efficiencies of 35.4\% for 
\(B^+ \to \mu^+ \nu_\mu\), 0.8\% for \(B \to X_u \ell^+ \nu_\ell\), and \(0.00011\)\% for continuum processes. The efficiency numbers here are defined as the fraction of selected events relative to the total number of generated events.

All eight categories, four for each experiment, are analyzed simultaneously in a combined likelihood fit with shared common systematic uncertainties, ensuring a consistent treatment of the two data sets.

\section{Validation using $B^+ \to \bar D^0 \pi^+$ decays}\label{re:CC_text}

To validate the modeling of the input variables used in the multivariate selection, we study the decay \(B^+ \to \bar{D}^0 \pi^+\) with \(\bar{D}^0 \to K^+ \pi^-\). The reconstruction procedure and selection criteria for the control channel in the Belle data remain as described in Ref.~\cite{Prim_2020}. The following outlines the strategy adopted for the Belle II analysis.

The charged particles from \(B^+ \to \bar{D}^0 \pi^+\) are required to satisfy the same impact-parameter criteria as the muon in $B^+ \to \mu^+ \nu_\mu$. The kaon and pion used to reconstruct the \(\bar{D}^0\) meson must each have a momentum greater than \SI{0.3}{GeV} in the c.m.\ frame. We require \(\mathcal{L}_K / (\mathcal{L}_K + \mathcal{L}_\pi) > 0.6\) and \(\mathcal{L}_\pi / (\mathcal{L}_K + \mathcal{L}_\pi) > 0.6\) for the kaon and pion, respectively. Only \(\bar{D}^0\) candidates with reconstructed masses within \SI{50}{MeV} of the expected value~\cite{PDG_2024} are retained. The $B^+ \to \bar D^0 \pi^+$ candidates are then reconstructed by combining the $\bar{D}^0$ candidates with a pion having momentum greater than \SI{2.1}{GeV} in the c.m.\ frame and \(\mathcal{L}_\pi / (\mathcal{L}_K + \mathcal{L}_\pi) > 0.6\).  
We require the reconstructed \(B\) meson mass to be within \SI{50}{MeV} of the expected value~\cite{PDG_2024}, a beam-constrained mass \(m_{\mathrm{bc}}\) above \SI{5.2}{GeV}, and an absolute energy difference \(|\Delta E|\) below \SI{0.2}{GeV}. The ROE reconstruction and selection for the control channel is identical to the signal channel described in the previous section. To further suppress continuum background, only events with \(\cos\theta_{\text{thrust}} < 0.8\) are retained, where \(\theta_{\text{thrust}}\) is defined as the angle between the thrust axis of the ROE and the thrust axis of the reconstructed \(B\) meson, in the c.m.\ frame.

\begin{figure}[htbp]
    \centering
    \includegraphics[width=\linewidth]{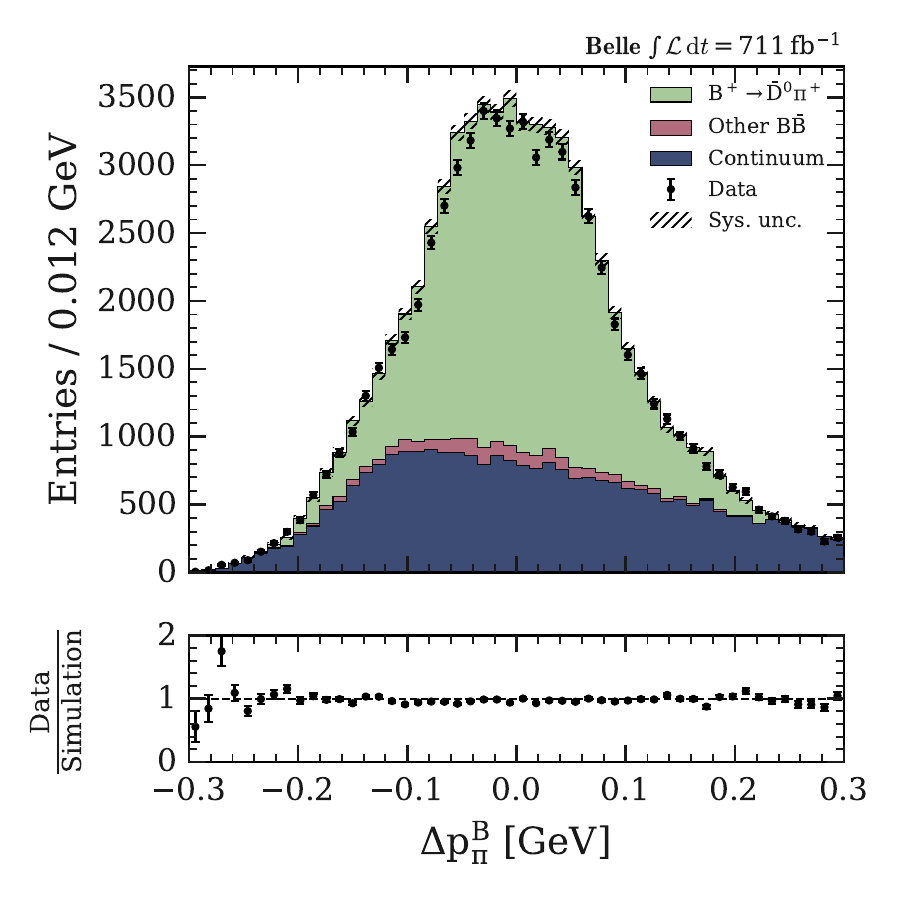}
    \includegraphics[width=\linewidth]{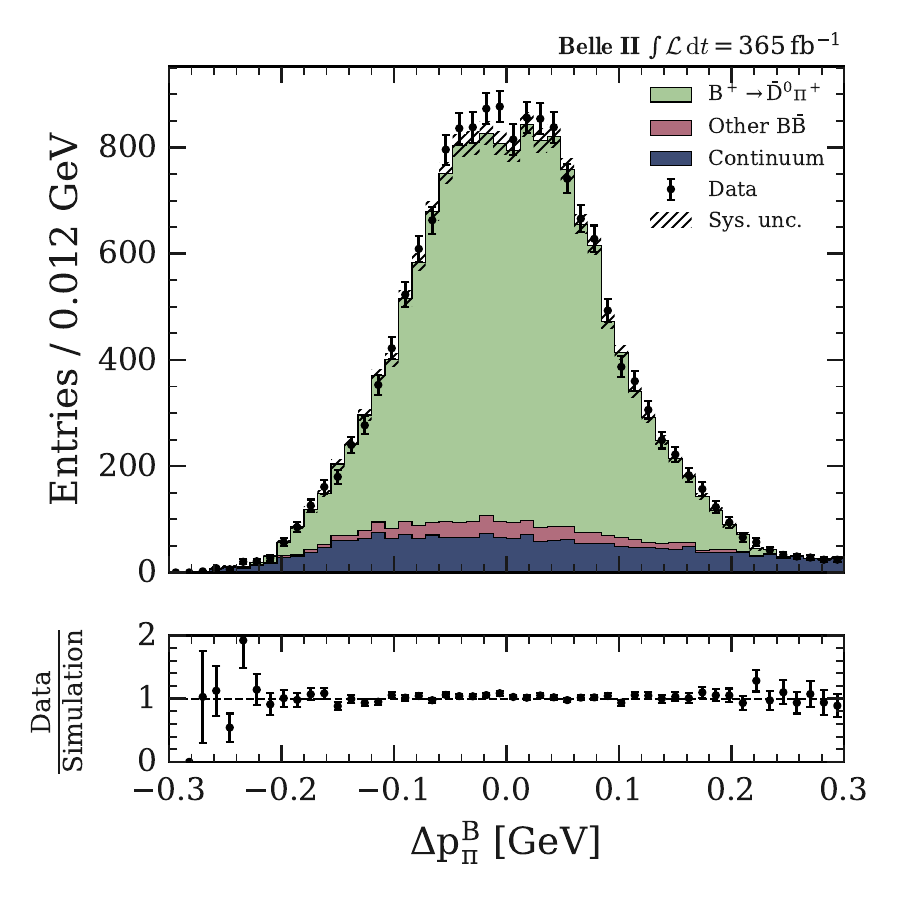}
    \caption{Upper panels: Distributions of the pion momentum bias in the $B$ meson rest frame for the $B^+ \to \bar D^0 \pi^+$ control channel, shown for Belle (top) and Belle~II (bottom) data sets and simulations. The colored stacked histograms show the different components: $B^+ \to \bar{D}^0[\to K^+ \pi^-] \pi^+$ in light green, other \(B \bar B\) decays in rose, and continuum events in dark blue. Data points are shown as black markers, while systematic uncertainties are indicated by black hatched bands. Lower panels: Ratios of data to simulation are shown, with error bars representing the combined statistical uncertainty of the data and systematic uncertainty of the simulation.}
    \label{fig:CC_Reso_Belle}
\end{figure}

The momentum of the pion originating from the \( B \) decay is boosted into the \( B \) rest frame using the reconstructed and calibrated ROE, and is then compared to the pion momentum in the same frame obtained using the fully reconstructed \( B \) meson. This comparison serves to identify potential biases and assess resolution differences between the data and simulation. The mean and standard deviation of the momentum bias, defined as 
\begin{equation}
    \Delta p_\pi^{B} = p_\pi^{B_{\text{ROE}}} - p_\pi^{B_\text{sig}},
\end{equation}
are determined and listed in Table~\ref{tab:resolutions_cc}. The \(\Delta p_\pi^{B}\) distributions from the Belle and Belle II data are shown in Fig. \ref{fig:CC_Reso_Belle}. The systematic uncertainties on the simulation arise from the uncertainty in the measured $B^+ \to \bar{D}^0[\to K^+ \pi^-] \pi^+$ branching fraction and limited MC sample size. We observe good agreement between simulation and the control sample data. 

\begin{table}[htbp]
\caption{Mean values and standard deviations for $B^+ \to \bar D^0 \pi^+$, each with their respective uncertainties, for the momentum bias of the data and simulation from the Belle and Belle~II experiments.}
\begin{tabular}{ l|c|c }
\hline \hline
Belle                        & $\mu$ & $\sigma$ \\ \hline
$\Delta p_\pi^{B}$ [GeV] MC   & $0.0119\pm0.0004$    & $0.1108\pm0.0003$          \\ 
$\Delta p_\pi^{B}$ [GeV] Data & $0.0123\pm0.0004$    & $0.1113\pm0.0003$          \\ \hline \hline
Belle~II                     & $\mu$ & $\sigma$ \\ \hline
$\Delta p_\pi^{B}$ [GeV] MC   & $0.0074\pm0.0004$    & $0.0942\pm0.0003$          \\ 
$\Delta p_\pi^{B}$ [GeV] Data & $0.0073\pm0.0007$    & $0.0932\pm0.0005$          \\ 
\hline \hline
\end{tabular}
\label{tab:resolutions_cc}
\end{table}

\begin{figure}[htbp]
    \centering
    \includegraphics[width=\linewidth]{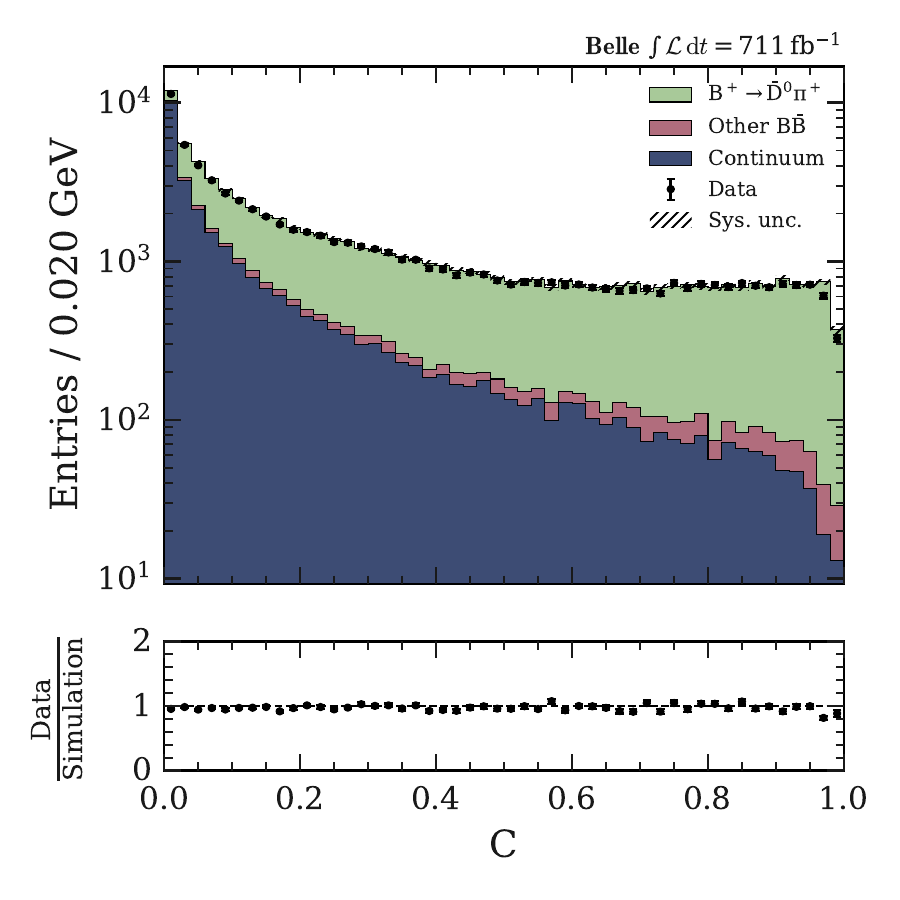}
    \includegraphics[width=\linewidth]{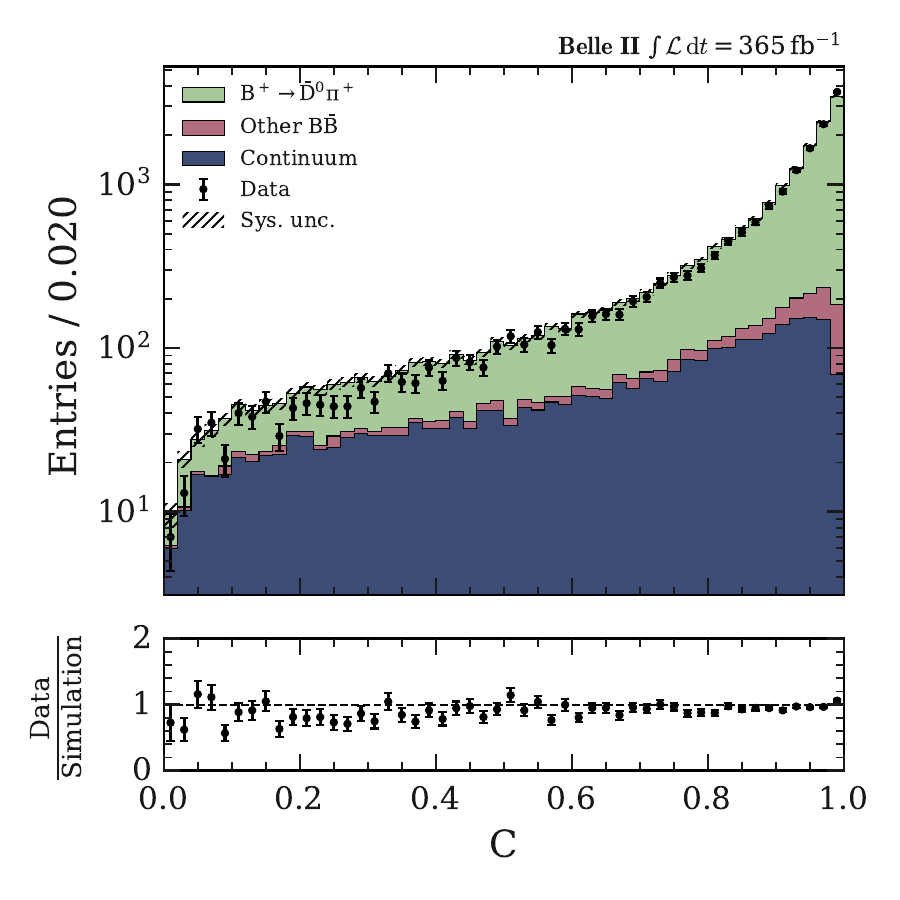}
    \caption{Distribution of the classifier output derived for $B^+ \to \mu^+ \nu_\mu$ and applied to the control $B^+ \to \bar D^0 \pi^+$ channel for the Belle (top) and Belle~II (bottom) data sets and simulation. For the Belle~II data the distribution is shown for the case where the $D$ meson is removed from the reconstruction and the pion momentum is scaled up. The color scheme for the individual components is the same as in Fig.~\ref{fig:CC_Reso_Belle}.}
    \label{fig:CC_BDT_Belle}
\end{figure}

To validate the modeling of the continuum-suppression classifier in Belle~II, an additional study is performed in which the reconstructed \(\bar{D}^0\) is omitted from the reconstruction to mimic a neutrino in the final state. Further, the momentum of the pion from the \(B\) decay is scaled to match the momentum of the muon in $B^+ \to \mu^+ \nu_\mu$. After these corrections, the sixteen event shape variables are recomputed and the classifier output $C$ is determined using the respective weights derived in the signal channel. The classifier is then used to study the selection efficiency in data and simulation for the Belle and Belle II categories. The distributions of the classifier output of the control channel for the Belle and Belle~II data sets are shown in Fig.~\ref{fig:CC_BDT_Belle}.

The data set is separated into the eight categories used in the $B^+ \to \mu^+ \nu_\mu$ analysis. The selection efficiency for the categories is summarized in Table~\ref{tab:sel_eff} for both data and simulation. The efficiency is defined as the fraction of reconstructed $B^+ \to \bar D^0 \pi^+$ events in a given region relative to the total number of reconstructed $B^+ \to \bar D^0 \pi^+$ events. The yield of $B^+ \to \bar D^0 \pi^+$ events in data is determined by a two-dimensional binned fit to the energy difference $\Delta E$ and the signal-side beam-constrained mass $m_{\mathrm{bc}}$. 

We perform \(\chi^2\) tests of the efficiency ratios between data and simulation for the four Belle and Belle~II categories separately and observe good agreement. For Belle, the ratios are compatible with unity within at most 
1.3 standard deviations, with a combined \(\chi^2 = 3.26\) corresponding to a \(p\)-value of 51\%. 
For Belle~II, the ratios agree with unity within at most 2.3 standard deviations, and the combined \(\chi^2 = 8.16\) yields a \(p\)-value of 9\%. Hence, no correction to the $B^+ \to \mu^+ \nu_\mu$\ efficiencies is applied. The efficiency ratios of data-to-simulation, together with the combined factors, are summarized in Table~\ref{CC_data_MC_ratio}.

\begin{table}[htbp]
\caption{Selection efficiencies of $B^+ \to \bar D^0 \pi^+$ and their uncertainties in eight categories for data and simulation from the Belle and Belle II experiments.}
\centering
\begin{tabular}{l|c|c}
\hline \hline
Belle & $\epsilon_{\text{Data}}$ & $\epsilon_{\text{MC}}$ \\ \hline
Category I   & $( 0.37\pm0.04 )\%$     & $( 0.43\pm0.04 )\%$   \\ 
Category II   & $( 0.35\pm0.04 )\%$     & $( 0.37\pm0.03 )\%$   \\ 
Category III   & $( 1.80\pm0.07 )\%$     & $( 1.92\pm0.09 )\%$   \\ 
Category IV   & $( 1.97\pm0.08 )\%$     & $( 2.04\pm0.10 )\%$   \\
\hline
\hline
Belle II & $\epsilon_{\text{Data}}$ & $\epsilon_{\text{MC}}$ \\ 
\hline
Category V   & $ \phantom{0}(7.25\pm0.26 )\%$     & $ \phantom{0}( 6.63\pm0.24 )\%$   \\ 
Category VI   & $( 11.99\pm0.33 )\%$     & $( 10.82\pm0.38 )\%$   \\ 
Category VII   & $( 16.99\pm0.40 )\%$     & $( 16.69\pm0.57 )\%$   \\ 
Category VIII  & $( 29.51\pm0.56 )\%$     & $( 30.23\pm1.02 )\%$   \\ 
\hline \hline
\end{tabular}
\label{tab:sel_eff}
\end{table}

\begin{table}[htbp]
	\caption{Ratios of data to simulation efficiencies for $B^+ \to \bar D^0 \pi^+$ and the combined factors in the eight categories for the Belle and Belle II experiments.}
	\begin{tabular}{l|c|c|c}
		\hline \hline
		Belle & $\epsilon_{\text{Data}} / \epsilon_{\text{MC}}$ & \( \chi^2 \) & $p$-value \\ \hline
		Category I   & $ 0.846\pm0.120 $  & \multirow{4}{*}{$3.26$} &  \multirow{4}{*}{$0.51$} \\ 
		Category II  & $ 0.950\pm0.135 $  & & \\ 
		Category III & $ 0.937\pm0.059 $  & & \\ 
		Category IV  & $ 0.967\pm0.059 $ & & \\ 
        \hline
        \hline
		Belle~II & $\epsilon_{\text{Data}} / \epsilon_{\text{MC}}$ & \( \chi^2 \) & $p$-value \\ \hline
		Category V  & $ 1.093\pm0.056$ & \multirow{4}{*}{$8.16$} &  \multirow{4}{*}{$0.09$} \\ 
		Category VI & $ 1.109\pm0.049 $ & &  \\ 
		Category VII  & $ 1.017\pm0.042 $  & & \\ 
		Category VIII   & $ 0.976\pm0.038 $  & & \\ 
        \hline \hline
	\end{tabular}
	\label{CC_data_MC_ratio}
\end{table}

\section{Statistical analysis and limit setting procedure}\label{Fit_setup}

To avoid experimenter’s bias, all selection criteria, reconstruction procedures, and fit models are defined and validated using simulation and control samples before inspecting the data in the signal region. The templates for signal and background momentum distributions are derived from simulation and use \num{22} uniform bins with a width of 50 MeV, covering the range \( p_\mu^B \in [2.2, 3.3] \)~GeV. 
A binned maximum-likelihood fit to the muon momentum \(p_\mu^B\) is performed to extract the branching fraction \(\mathcal{B}(B^+ \to \mu^+ \nu)\). Systematic uncertainties are incorporated with multiplicative or additive event-count modifiers in the likelihood.

The likelihood ${\cal L}$ is constructed as the product of the likelihoods \(\mathcal{L}_c\) for each category $c$ and the Gaussian constraints \(\mathcal{G}_k\) for each process $k$:
\begin{align} \label{eq:likelihood}
 {\cal L} = \prod_c {\cal L}_c \times \prod_k \mathcal{G}_k \, .
\end{align}
The likelihood of each category is the product of Poisson probability density functions \(\mathcal{P}\) over bins $i$ of the muon momentum spectrum,
\begin{align}
    {\cal L}_c = \prod_{i}^{\mathrm{bins}} \mathcal{P}\left(n_i | \nu_i \right) \, ,
\end{align}
with \(n_i\) and \(\nu_i\) denoting the number of observed and expected events. The expected number of events in each bin, \( \nu_i \), is determined from simulation and is 
given by 
\begin{equation}\label{eq:nui}
 \nu_i = \sum_k^{\rm processes} \, f_{ik} \, \eta_k \, ,
\end{equation}
where \( \eta_k \) represents the total number of events from process \( k \), and \( f_{ik} \) denotes the fraction of these events reconstructed in bin \( i \).

Systematic uncertainties are incorporated into the likelihood through a vector of nuisance parameters (NPs), \(\boldsymbol{\theta}_n\), which parameterize both normalization and shape variations. For normalization uncertainties, each independent source $s$ is modeled by a Gaussian-constrained parameter. Its effect is incorporated by modifying the total yield of process $k$ as 
\begin{equation}
    \eta_k \to \eta_k \left( 1 + \theta_{ks} \right) \,,
\end{equation}
where \(\theta_{ks}\) denotes the NP parameterizing the corresponding uncertainty. 

For the shape uncertainties, each element of the nuisance parameter vector \(\boldsymbol{\theta}_n\) corresponds to a bin in the fitted \( p_\mu^B \) spectrum across the eight event categories of a given template. The NPs are constrained in the likelihood function through multivariate Gaussian distributions, 
\begin{align}
    \mathcal{G}_k = \mathcal{G}_k( \boldsymbol{0}; \boldsymbol{\theta}_k, \Sigma_k ) \,,
\end{align}
where \( \Sigma_k \) represents the systematic covariance matrix associated with template \( k \). This matrix contains all sources of systematic uncertainty that affect template \( k \) and is given by 
\begin{align}
    \Sigma_k = \sum_{s}^{\text{error sources}} \Sigma_{ks} \,,
\end{align}
where \( \Sigma_{ks} \) is the contribution of a specific uncertainty source \( s \). 
The construction of \( \Sigma_{ks} \) is based on the uncertainty vector \( \boldsymbol{\sigma_{ks}} \), whose elements correspond to the absolute uncertainties in the \( p_\mu^B \) bins for template \( k \). 

Systematic uncertainties originating from the same source are treated either as fully correlated, 
\begin{equation}
    \Sigma_{ks} = \boldsymbol{\sigma_{ks}} \otimes \boldsymbol{\sigma_{ks}} \,,
\end{equation}
or as uncorrelated, as i.e. in the case of the statistical uncertainties of the simulation, 
\begin{align}
    \Sigma_{ks} = \text{Diag}\left( \boldsymbol{\sigma_{ks}}^2 \right) \,.
\end{align}  

The NPs are incorporated into Eq.~\ref{eq:nui} by modifying the fractions \( f_{ik} \) for all templates. Specifically, to account for shape uncertainties, this leads to
\begin{equation}
    f_{ik} = \frac{ \eta_{ik} }{ \sum_j \eta_{jk} } \to  
    \frac{ \eta_{ik} \left( 1 + \theta_{ik} \right) }{ \sum_j \eta_{jk} \left( 1 + \theta_{jk} \right)  }.
\end{equation}
Here $\eta_{ik}$ denotes the expected yield of process $k$ in bin $i$. 

The fit contains several templates, five of which have their normalizations treated as free parameters: the $B^+ \to \mu^+ \nu_\mu$\ signal, semileptonic \(b \to u\) and \(b \to c\) transitions, and the continuum backgrounds for the Belle and Belle~II data sets. The continuum background contributions are corrected using data-driven methods specific to each experiment. Although the underlying processes are the same, the observed distributions differ between Belle and Belle~II owing to detector responses, selection efficiencies, and acceptances. Therefore, the continuum backgrounds are modeled separately and not combined in the fit. The yields of the background contributions from \(B^+ \to \mu^+ \nu_\mu \gamma\) and \(b \to s\) transitions are constrained to the expected branching fractions within the corresponding uncertainties, since they resemble the signal distribution.

The likelihood in Eq.~\ref{eq:likelihood} is maximized using the \texttt{iMinuit} package~\cite{iminuit} to determine the yields \(\eta_k\) of the fit components. The numbers of signal and semileptonic \(b \to u\) events are translated into partial or full branching fractions using the corresponding efficiencies and the numbers of recorded \(B\) meson pairs in the Belle and Belle~II data sets. The fit accounts for \(8 \times 154\) shape uncertainties and 73 additional NPs associated with normalization uncertainties. The fit procedure is validated using ensembles of pseudo-data sets for different input $B^+ \to \mu^+ \nu_\mu$ branching fractions. No biases in the extracted central value and uncertainty coverage are observed. 

We also investigate sterile neutrino and WA contributions by introducing additional free parameters into the fit. When setting limits on sterile neutrinos, the $B^+ \to \mu^+ \nu_\mu$\ yield is constrained to its SM expectation. This is discussed in more detail in Sections \ref{ref:results} and \ref{ref:wa_proc}.

We construct confidence intervals for the processes using the profile likelihood ratio method. For a given process $\eta_k$, the likelihood ratio test statistic is defined as 
\begin{equation} \label{eq:test_stat}
  \Lambda(\eta_k) =  - 2 \ln \frac{ \mathcal{L}( \eta_k, \boldsymbol{\widehat \eta_{\eta_k}}, \boldsymbol{\widehat \theta_{\eta_k}}  ) }{  \mathcal{L}( \widehat  \eta_k,  \boldsymbol{ \widehat \eta}, \boldsymbol{ \widehat \theta}  )  } \, ,
\end{equation}
where $ \widehat{\eta}_k$, $\boldsymbol{\widehat \eta}$, and $\boldsymbol{ \widehat \theta} $ correspond to the values of the process of interest, the remaining processes, and a vector of NPs that maximize the likelihood function unconditionally. In contrast, $\boldsymbol{\widehat \eta_{\eta_k}}$ and $\boldsymbol{\widehat \theta_{\eta_k}}$ denote the values of the remaining processes and NPs that maximize the likelihood under the constraint that the process of interest is fixed at a given value $\eta_k$. In the asymptotic limit, the test statistic in Eq.~\ref{eq:test_stat} can be used to construct approximate confidence intervals at confidence level $\alpha_{\text{CL}}$ according to
\begin{equation}
  1 - \alpha_{\text{CL}} = \int_{ \Lambda(\eta_k) }^{\infty} \, f_{\chi^2}(x; 1) \, \text{d} x \, ,
\end{equation}
where $f_{\chi^2}(x; 1)$ represents the $\chi^2$ distribution with a single degree of freedom. 

In the absence of a significant signal, we determine both frequentist and Bayesian limits. For the frequentist one-sided (positive) limit, we modify our test statistic following the prescription of Refs.~\cite{Aad_2012,Cowan_2011} to 
\begin{equation}\label{eq:imp_test_stat}
\begin{aligned}
  q_0(\eta_k) = \begin{cases} 
    \Lambda(\eta_k) & \quad \text{if } \eta_k \geq 0 \, , \\
    - \Lambda(\eta_k)  & \quad \text{if } \eta_k < 0 \, , 
  \end{cases}  
\end{aligned}
\end{equation}
which enhances the sensitivity of our analysis. This test statistic is asymptotically distributed as 
\begin{equation}
 f(q_0) = \frac{1}{2}  f_{\chi^2}(-q_0; 1) + \frac{1}{2}  f_{\chi^2}(q_0; 1) \, .
\end{equation}
Given an observed test statistic value $q_0^{\rm obs}$, the (local) probability of observing a signal, $p_0$, is computed as 
\begin{equation}\label{eq:loc_prob}
 p_0 = \int_{q_0^{\rm obs}}^\infty \, f(q_0) \, \text{d} q_0 \, .
\end{equation}

\section{Systematic uncertainties}\label{ref:sys_uncerts}

Two types of systematic uncertainties are included, additive and multiplicative.

Additive uncertainties affect the shape of the distributions. Several sources of systematic uncertainty affect the measurement of $B^+ \to \mu^+ \nu_\mu$, the largest arising from the modeling of the \(b \to u\) background. Since the overall normalization of these decays is determined from data, only shape uncertainties need to be considered. These originate from uncertainties in the form-factor parameters of the \(B \to \pi \ell^+ \nu_\ell\), \(B \to \rho \ell^+ \nu_\ell\), \(B^+ \to \omega \ell^+ \nu_\ell\), \(B^+ \to \eta \ell^+ \nu_\ell\), and \(B^+ \to \eta' \ell^+ \nu_\ell\) decays; from the measured branching fractions of these modes; and from the modeling of the non-resonant component of the \(b \to u\) background. The DFN model used to describe the non-resonant transitions is parameterized by the \(b\)-quark mass and a non-perturbative parameter, whose uncertainties propagate to the template shape. An alternative description using the Bosch--Lange--Neubert--Paz (BLNP) model~\cite{Lange:2005yw} is also considered, and the difference between the DFN and BLNP predictions is taken as an additional systematic uncertainty.

The second-largest systematic uncertainty arises from the modeling of the continuum background. The off-resonance data used to correct the continuum contribution are subject to statistical uncertainties from the finite sample size, which are evaluated using a bootstrapping procedure that generates multiple replicas by resampling the data set.

Uncertainties in \(b \to c\) transitions, dominated by \(B \to D \ell^+ \nu_\ell\) and \(B \to D^* \ell^+ \nu_\ell\) decays, also affect the measurement. These arise from the form-factor parameters and their branching fractions, which are propagated into the model. In addition, an uncertainty is included to account for the branching fraction of hadronic \(B \to D \pi\) decays, which contribute near the kinematic endpoint.

The rare background contribution is dominated by \(B^+ \to K^0 \pi^+\) and \(B^0 \to K_0^{*+} \pi^-\) decays, for which 
uncertainties in the measured branching fractions are included as systematic uncertainties. The uncertainty on the \(B^+ \to \mu^+ \nu_\mu \gamma\) background is evaluated using both experimental and theoretical inputs. The experimental component reflects the precision of the measured branching fraction, while the theoretical component accounts for variations in model parameters. These include the ratio of the inverse moment to the zeroth moment of the light-cone wave function, which describes the longitudinal momentum distribution of the light quark inside the \(B\) meson, the \(b\)-quark mass \(m_b\), and the assumption of whether the vector and axial form factors are taken to be equal~\cite{Korchemsky:1999qb}. The model is varied by adjusting these parameters, and the largest deviation from the nominal prediction is assigned as a systematic uncertainty.

The $B^+ \to \mu^+ \nu_\mu$ signal shape is affected by uncertainties in the particle-identification efficiencies between data and simulation, and the finite size of the simulated sample, which are included as additive uncertainties.

Multiplicative uncertainties scale the overall normalization and shift the extracted branching fraction.

The efficiency ratios between data and simulation, obtained from the validation study in Section~\ref{re:CC_text}, are used to evaluate the systematic uncertainty on the efficiency.  For Belle and Belle~II, the averaged ratios are \(0.941 \pm 0.038\) and \(1.034 \pm 0.022\), respectively. The quoted uncertainties on these averages are used as the systematic uncertainties on the efficiency.

For the $B^+ \to \mu^+ \nu_\mu$ signal, additional systematic uncertainties originate from tracking, differences in particle-identification efficiencies between data and simulation, and the uncertainty in the total number of \(B\bar{B}\) pairs. These uncertainties are treated as normalization uncertainties.

Finally, for all distributions considered, the impact of the finite simulation sample size is included.

Table~\ref{table:sys_table_comb} provides a summary of the systematic uncertainties affecting the $B^+ \to \mu^+ \nu_\mu$\ branching fraction measurement.

\begin{table}[htbp]
	\caption{The uncertainties on the measured $B^+ \to \mu^+ \nu_\mu$ branching fraction are shown. For definitions of additive and multiplicative errors see the text. FF denotes form factor, and BF denotes branching fraction.}
\begin{tabular}{l|c}
    \hline \hline
	Source                                & Fractional \\ 
	     & uncertainty \\ \hline \hline 
    \textbf{Additive uncertainties}&\\ \hline
        $\quad  $$b \to u$ modeling&17.4\%\\ 
	$\quad \quad $$B \to \pi \ell^+ \nu_\ell$ FF&8.2\%\\ 
	$\quad \quad $$B \to \rho \ell^+ \nu_\ell$ FF&8.9\%\\ 
	$\quad \quad $$B^+ \to \omega \ell^+ \nu_\ell$ FF&4.5\%\\ 
	$\quad \quad $$B^+ \to \eta \ell^+ \nu_\ell$ FF&0.2\%\\ 
	$\quad \quad $$B^+ \to \eta' \ell^+ \nu_\ell$ FF&1.0\%\\ 
	$\quad \quad $$B \to \pi \ell^+ \nu_\ell$ BF&4.8\%\\ 
	$\quad \quad $$B \to \rho \ell^+ \nu_\ell$ BF&1.0\%\\ 
	$\quad \quad $$B^+ \to \omega \ell^+ \nu_\ell$ BF&0.2\%\\ 
	$\quad \quad $$B^+ \to \eta \ell^+ \nu_\ell$ BF&0.2\%\\ 
	$\quad \quad $$B^+ \to \eta' \ell^+ \nu_\ell$ BF&0.1\%\\ 
	$\quad \quad $$B \to X_u \ell^+ \nu_\ell$ BF&3.3\%\\ 
	$\quad \quad $DFN parameters&5.7\%\\ 
	$\quad \quad $Hybrid model&6.0\%\\ 
	$\quad \quad $MC sample size&5.6\%\\ 
    
	$\quad $Continuum modeling&14.0\%\\ 
	$\quad \quad$Shape correction&4.1\%\\ 
	$\quad \quad$MC sample size&13.3\%\\

	$\quad $$B^+ \to \mu^+ \nu_\mu \gamma$ modeling&4.8\%\\     

	$\quad $$b \to c$ modeling&2.4\%\\ 
	$\quad $Rare decay modeling&1.5\%\\ 

	$\quad B^+ \to \mu^+ \nu_\mu$ modeling&1.5\%\\  \hline \hline 
    \textbf{Multiplicative uncertainties}&\\  \hline  
    $\quad $PID efficiency Belle &1.6\%\\ 
    $\quad $Tracking efficiency Belle &0.3\%\\ 
    $\quad $\(N_{B\bar{B}}\) Belle &1.1\%\\ 
    $\quad $PID efficiency Belle II&1.0\%\\ 
    $\quad $Tracking efficiency Belle II&0.1\%\\ 
    $\quad $\(N_{B\bar{B}}\) Belle II&0.3\%\\ 
    $\quad $$f_{+0}$&2.1\%\\ 
    $\quad $Control channel efficiency ratios & 3.2\%\\ \hline \hline 
	\textbf{Total systematic uncertainty}&23.6\%\\ 
	\textbf{Total statistical uncertainty}&42.9\%\\ 
    \hline \hline
\end{tabular}
    \label{table:sys_table_comb}
\end{table}

Covariance matrices for all categories and $p_\mu^B$ bin are generated independently for each eigendirection of the Belle and Belle~II data sets. Systematic uncertainties common to both experiments, such as branching fractions, are treated as fully correlated, whereas uncertainties intrinsic to each experiment, such as the number of \(B\bar{B}\) pairs, are treated as uncorrelated.

Assuming a branching fraction of \(\mathcal{B}(B^+ \to \mu^+ \nu_\mu) = (4.18 \pm 0.44) \times 10^{-7}\), and 
including both statistical and systematic uncertainties, the expected one-sided signal significance is \(2.3^{+0.7}_{-0.8}\) standard deviations, as estimated with an Asimov data set~\cite{Cowan:2010js}.

\section{Results}\label{ref:results}

\subsection{Search for $B^+ \to \mu^+ \nu_\mu$}

\begin{figure}[htbp]
    \centering
    \includegraphics[width=\linewidth]{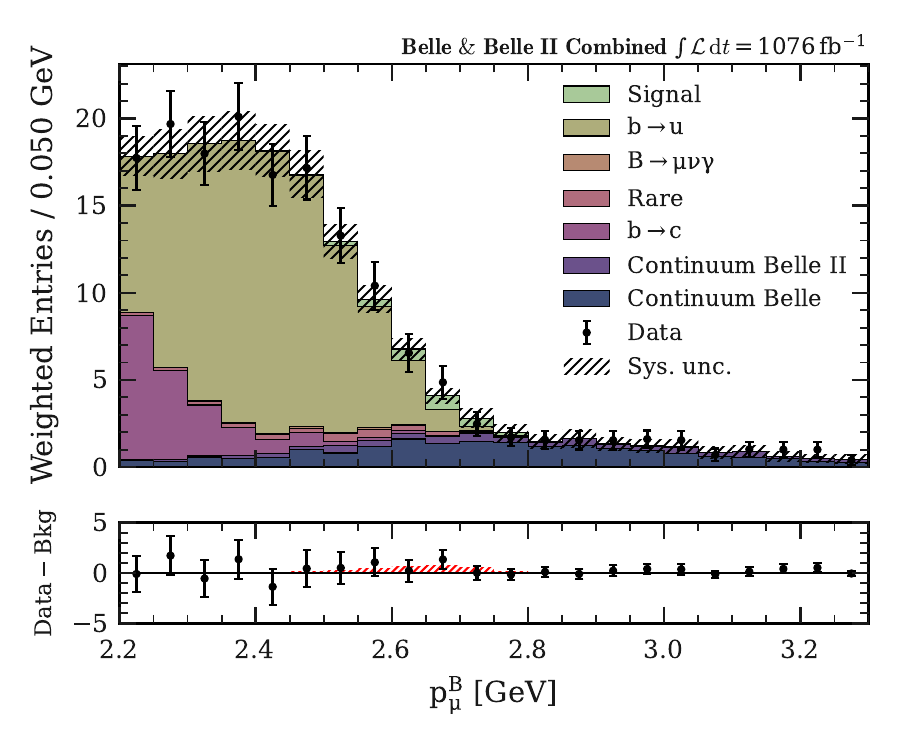}
    \caption{
    The upper panel: Weighted distribution of the fitted muon momentum \(p_\mu^B\) in the \(B\) rest frame, where events from the signal-enriched category I and V are weighted by $w = \log \left( 1 + S_c / B_c \right)$. The stacked histograms represent the fitted signal and background processes with the color scheme introduced in Fig.~\ref{fig:BDT_Output_B2}, and the data points show the combined Belle and Belle II sample. The hatched band indicates the systematic uncertainty. The lower panel: Difference between data and the background prediction, with uncertainties from statistical and systematic sources. The red hatched area shows the fitted signal process.}
    \label{fig:weighted_fit}
\end{figure}

Figure~\ref{fig:weighted_fit} presents the combined muon-momentum distribution in the $B$ meson rest frame, where events from the signal-enriched Categories~I and~V are weighted by \mbox{$\log \left( 1 + S_c / B_c \right)$}~\cite{ATLAS:2012yve}, with $S_c$ and $B_c$ denoting the numbers of signal and background events, respectively, within a 68\% containment window centered on the $B^+ \to \mu^+ \nu_\mu$ peak. This weighting enhances the sensitivity to regions with a higher signal-to-background ratio; distributions for the eight classifier categories can be found in Figure~\ref{fig:fit_eight_cats} in Appendix~\ref{sec:appendix_A}. We assume no contributions from WA processes and set these contributions to zero. The overall fit quality is good, with a post-fit $p$-value of 90\% obtained from a $\chi^2$ test. 

The branching fraction is measured to be 
\begin{align}\label{eq:final_bf}
    \mathcal{B}(B^+ \to \mu^+ \nu_\mu) = (4.4 \pm 1.9 \pm 1.0) \times 10^{-7} \, ,
\end{align}
where the first uncertainty is statistical and the second systematic. This corresponds to an observed signal yield of \(\nu^{B^+ \to \mu^+ \nu_\mu} = 150 \pm 73\) events. This represents the most precise determination of the $B^+ \to \mu^+ \nu_\mu$ branching fraction to date, benefiting from improved modeling of the background shapes and a larger data sample by adding the Belle II data. The systematic uncertainty is marginally larger than in the previous Belle result Ref.~\cite{Prim_2020}, owing to updated modeling of the \(b \to u\) and \(b \to c\) backgrounds and the inclusion of an additional efficiency uncertainty from the control mode. The observed significance relative to the background-only hypothesis is 2.4 standard deviations, consistent with the expectation of $2.3^{+0.7}_{-0.8}$.

\begin{figure}[htbp]
    \centering
    \includegraphics[width=\linewidth]{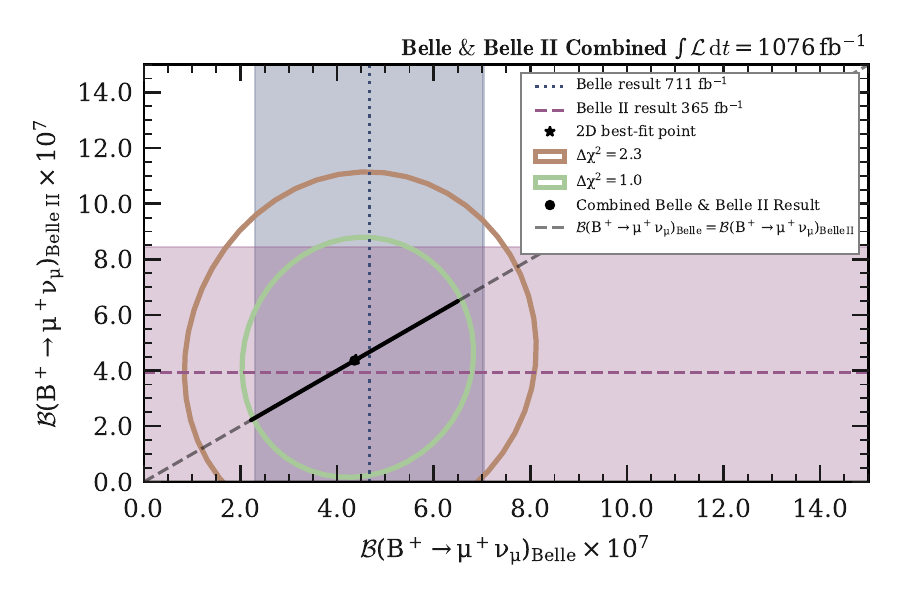}
    \caption{Two-dimensional likelihood contours for the $B^+ \to \mu^+ \nu_\mu$ branching fraction measured from combined Belle and Belle II data. The inner light green and outer golden brown ellipses correspond to the joint 38.3\% (\(\Delta\chi^2 = 1.0\)) and 68.3\% (\(\Delta\chi^2 = 2.3\)) confidence regions, respectively. The dashed dark blue vertical line with its shaded band shows the Belle standalone result (Eq.~\ref{eq:final_bf_B}), while the dashed rose horizontal line with its shaded band shows the Belle~II standalone result (Eq.~\ref{eq:final_bf_B2}). The black star marks the best-fit point when Belle and Belle II branching fractions are fitted independently, and the black circle indicates the combined fit with the single branching fraction extracted across both data sets, as shown in Eq.~\ref{eq:final_bf}.}
    \label{fig:Independent_results}
\end{figure}

In addition to the nominal fit combining signal and background across Belle and Belle~II, we perform the fit with the background combined but the signal treated independently, which yields a consistent result. We also carry out fully standalone fits for Belle and Belle~II with neither signal nor background combined. This is shown in Fig.~\ref{fig:Independent_results} in addition to the combined Belle and Belle~II result. The standalone fits yield
\begin{align}\label{eq:final_bf_B}
\mathcal{B}(B^+ \to \mu^+ \nu_\mu)_{\text{Belle}} &= (4.7 \pm 2.1 \pm 1.1) \times 10^{-7}, \\
\label{eq:final_bf_B2}
\mathcal{B}(B^+ \to \mu^+ \nu_\mu)_{\text{Belle~II}} &= (3.9 \pm 3.8 \pm 2.5) \times 10^{-7},
\end{align}
respectively. The shift in the Belle central value compared to the published result in Ref.~\cite{Prim_2020} arises from updated central values and systematic uncertainties in the \(b \to u\) and \(b \to c\) backgrounds. The updated Belle result supersedes the previous measurement. Although the overall uncertainty has increased slightly, the modeling of the \(b \to u\) contribution employs a more advanced description compared to the previous iteration of this analysis. While the DFN model is retained, the underlying scheme for its input parameters has been updated from the \(1S\) to the Kagan–Neubert scheme~\cite{Kagan:1998ym,Buchmuller:2005zv}. In addition, the branching fractions and form factors have been updated to the latest values in Refs.~\cite{PDG_2024,HFLAV_2024}, which are constrained by new auxiliary measurements, resulting in a more up-to-date representation of this contribution. The Belle result is consistent within 1.1 standard deviations with the previous one, where the comparison accounts for correlations by treating the statistical and most systematic uncertainties as correlated, while updated uncertainties from new theoretical inputs are treated as uncorrelated.

The magnitude of the CKM matrix element \(|V_{ub}|\) is extracted from the branching fraction in Eq.~\ref{eq:final_bf} using the \(B\) meson decay constant \(f_B = (190 \pm 1.3)\,\mathrm{MeV}\)~\cite{FLAG_2024} and the \(B^+\) lifetime \(\tau_B = (1.638 \pm 0.004)\,\mathrm{ps}\)~\cite{HFLAV_2024}. The result is
\begin{equation}
    |V_{ub}| = \left(3.92\,^{+\,0.77}_{-\,0.96}\,(\text{stat.})\,^{+\,0.44}_{-\,0.49}\,(\text{sys.})\,\pm\,0.03\,(\text{theo.})\right)\times 10^{-3} \, ,
\end{equation}
which, while less precise, is consistent with recent inclusive and exclusive determinations of \(|V_{ub}|\)~\cite{PDG_2024}.

\begin{figure}[htbp]
    \centering
    \includegraphics[width=\linewidth]{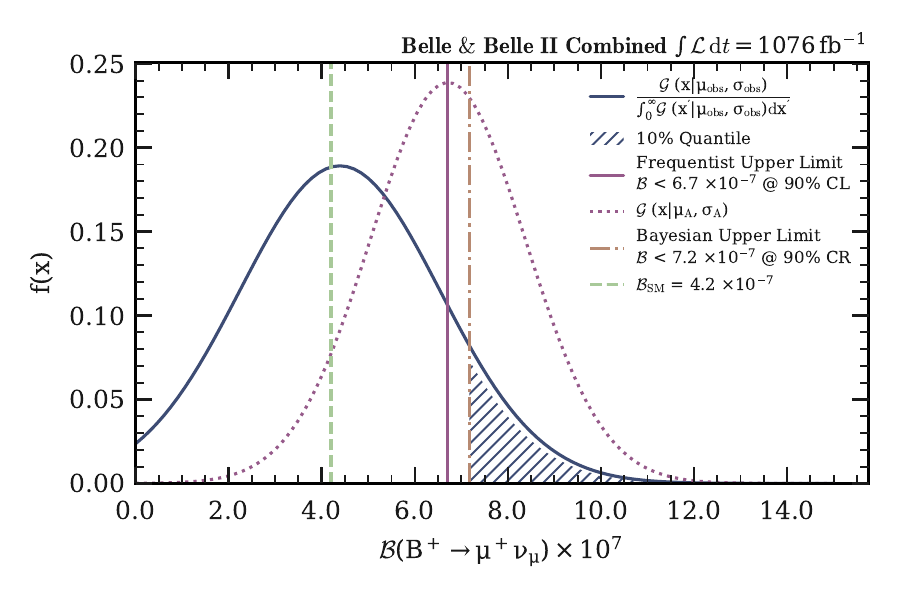}
    \caption{The observed Bayesian (golden brown dash-dotted line) upper limit at 90\% CrL and frequentist (medium purple solid line) upper limit at 90\% CL, along with the corresponding Bayesian (dark blue curve) and frequentist (medium purple dotted curve) PDFs. The SM expectation for the $B^+ \to \mu^+ \nu_\mu$\ branching fraction is indicated by the light green dashed line.}
    \label{fig:Upper_limits}
\end{figure}

Due to the low significance of the observed $B^+ \to \mu^+ \nu_\mu$ branching fraction, both Bayesian and Frequentist upper limits are determined. The likelihood function is converted into a Bayesian posterior probability density function (PDF) by assuming a flat prior, \(\pi(\eta_k)\), on the partial branching fraction, resulting in
\begin{equation}
   f(\eta_k | n ) = \frac{\mathcal{L}( n |\eta_k)\pi(\eta_k)}{\int_{0}^{\infty} \mathcal{L}( n |\eta_k)\pi(\eta_k)\,\mathrm{d}\eta_k}\, ,
\end{equation}
where \( n \) denotes the vector of the observed yields in the bins of all categories. The Bayesian upper limit at 90\% credibility level (CrL) is determined to be
\begin{align}
    \mathcal{B}(B^+ \to \mu^+ \nu_\mu) <  7.2 \times 10^{-7} \, .
\end{align}
Performing the fit to ensembles of Asimov data sets with the NPs shifted to their best fit values, we determined the frequentist upper limit 
\begin{align}
    \mathcal{B}(B^+ \to \mu^+ \nu_\mu) < 6.7 \times 10^{-7}
\end{align}
at  90\% confidence level (CL). Both upper limits are shown in Fig. \ref{fig:Upper_limits} in addition to the SM expectation. 

\subsection{2HDM Interpretation}

\begin{figure}[htbp]
    \centering
    \includegraphics[width=\linewidth]{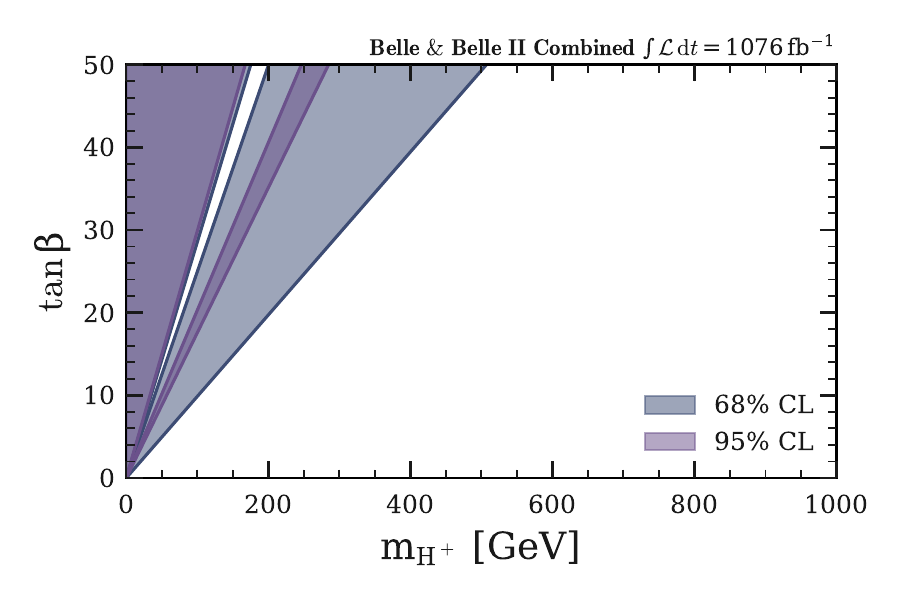}
        \includegraphics[width=\linewidth]{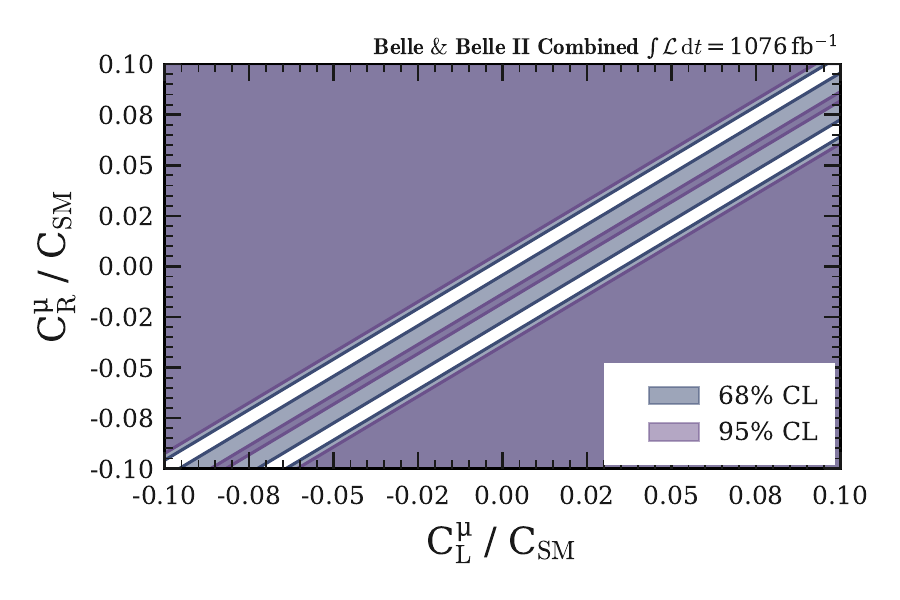}
    \caption{Excluded regions in the parameter space of the 2HDM of type II (top) and type III (bottom). The filled blue regions show the values excluded at the 68\% (dark blue) and 95\% CL (dark purple).}
    \label{fig:New_Physics_tanb_mh}
\end{figure}

Extending the study of the $B^+ \to \mu^+ \nu_\mu$ decay to physics beyond the SM, the measured branching fraction is used to constrain the parameter space of two-Higgs-doublet models (2HDM) of type II~\cite{Higgs_Type_2} and type III~\cite{Higgs_Type_2_a,Higgs_Type_2_b}, using the process shown in Fig.~\ref{fig:Feynmans}~(b-c). In type II models, the presence of a charged Higgs boson modifies the branching fraction to
\begin{equation}
    \mathcal{B}(B^+ \to \mu^+ \nu_\mu) = \mathcal{B}_{SM} \times \left( 1 - \frac{m_B^2 \tan^2 \beta}{m_{H^+}^2}\right) \, .
\end{equation}
Here, \(\mathcal{B}_{SM}\) is the SM branching fraction, \(\tan \beta\) is the ratio of the vacuum expectation values of the two Higgs fields, and \(m_{H^+}\) is the mass of the charged Higgs boson. In type III models, the branching fraction is modified as
\begin{equation}
    \mathcal{B}(B^+ \to \mu^+ \nu_\mu) = \mathcal{B}_{SM} \times \left| 1 + \frac{m_B^2 }{m_b m_\mu} \left( \frac{C_\text{R}^\mu}{C_\text{SM}} - \frac{C_\text{L}^\mu}{C_\text{SM}}\right)\right|^2 \, ,
\end{equation}
where \(m_b\) is the mass of the \(b\) quark, \(C_\text{SM}\) is the SM coupling and \(C_\text{R/L}^\mu\) are the Wilson coefficients that encode the beyond SM physics contribution. The excluded parameter regions for the 2HDM of type II and type III at 68\% and 95\% confidence levels are shown in Fig.~\ref{fig:New_Physics_tanb_mh}. These exclusion limits are obtained by constructing a $\chi^2$ statistic from the observed and predicted branching fractions, as defined in Eq.~(\ref{eq:final_bf}) and Eq.~(\ref{eq:munu_bf}), respectively. The limits presented here supersede those obtained in the previous Belle analysis~\cite{Prim_2020}.

A search for stable sterile neutrinos \(N\), shown in Fig.~\ref{fig:Feynmans} (c) and Fig.~\ref{fig:Feynmans} (d), is performed by fitting the muon momentum in the \(B\) meson rest frame, with the $B^+ \to \mu^+ \nu_\mu$ contribution fixed to its SM expectation. An additional template corresponding to \(B^+ \to \mu^+ N\) processes is included, assuming the same momentum distribution as for $B^+ \to \mu^+ \nu_\mu$. The fit is performed for selected neutrino mass \(m_N\) hypotheses in the range \SI{0.0}{\giga \eV} \(\leq m_N \leq \) \SI{1.5}{\giga \eV} with the momentum distribution of \(B^+ \to \mu^+ N\) shifted for each value to reflect the kinematic changes. The largest shift expected is \SI{0.2}{\giga \eV} at \(m_N = \SI{1.5}{\giga \eV}\).

We assume the sterile neutrino to be stable and to not decay inside the detector. Based on simulations, this assumption is valid for lifetimes \(\tau_N > 0.5\times10^{-6} \,\mathrm{s} \), where fewer than 5\% of sterile neutrinos decay before exiting the detector. If a sterile neutrino would decay inside the detector, its decay products would appear in the ROE. This would change the reconstructed boost to the \(B\) rest frame, causing events with and without neutrino decays to have different muon momentum distributions. Instead of a single peak in the boosted muon momentum \(p_\mu^B\), two distinct peaks would appear. To account for the small fraction of potentially decaying neutrinos at the threshold lifetime, an additional systematic uncertainty is assigned based on the percentage of sterile neutrinos that could decay within the detector~\cite{Asaka:Ishida:code}.

\begin{figure}[htbp]
    \centering
    \includegraphics[width=\linewidth]{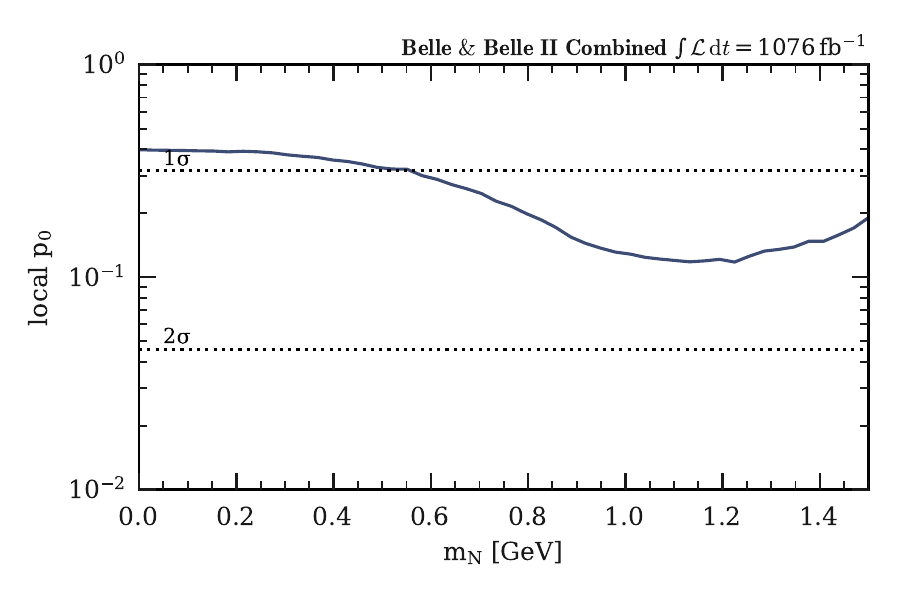}
    \includegraphics[width=\linewidth]{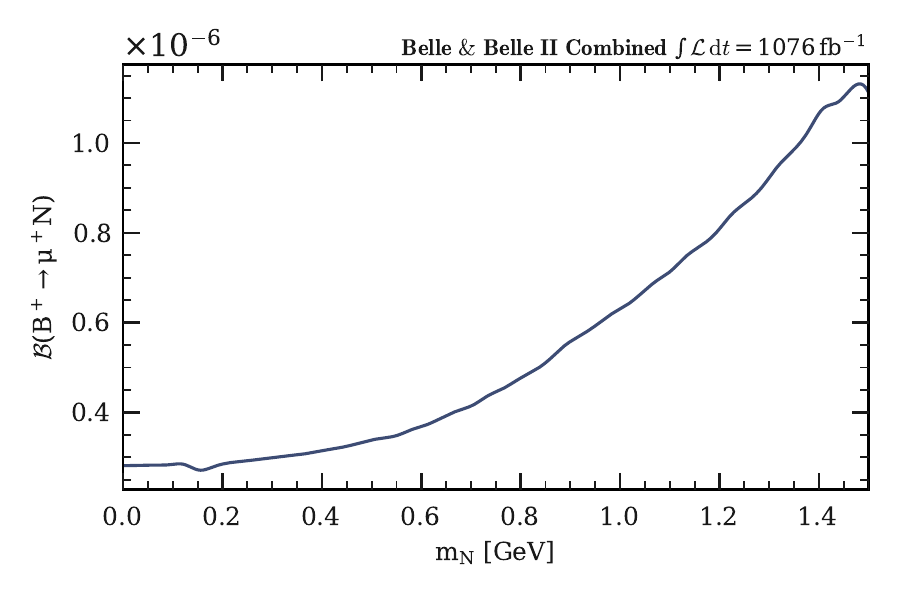}
    \caption{Local \(p_0\) value determined from the fit using an additional template for \(B^+ \to \mu^+ N\) and assuming the background-only hypothesis (top), and the Bayesian upper limit on the branching fraction of \(B^+ \to \mu^+ N\) at 90\% CrL (bottom).}
    \label{fig:Sterile_Ulocal_p}
\end{figure}

The significance over the background-only hypothesis is determined as described in Eq. \ref{eq:imp_test_stat} and shown in Fig. \ref{fig:Sterile_Ulocal_p}. There is no significant excess observed with the largest local significance over the background-only hypothesis of 1.6 standard deviations at a mass of around \(m_N = \) \SI{1.2}{\giga \eV}. The global significance, accounting for the look-elsewhere effect and estimated using toy MC data sets, is found to be 0.6 standard deviations at this mass.

\begin{figure}[htbp]
    \centering
    \includegraphics[width=\linewidth]{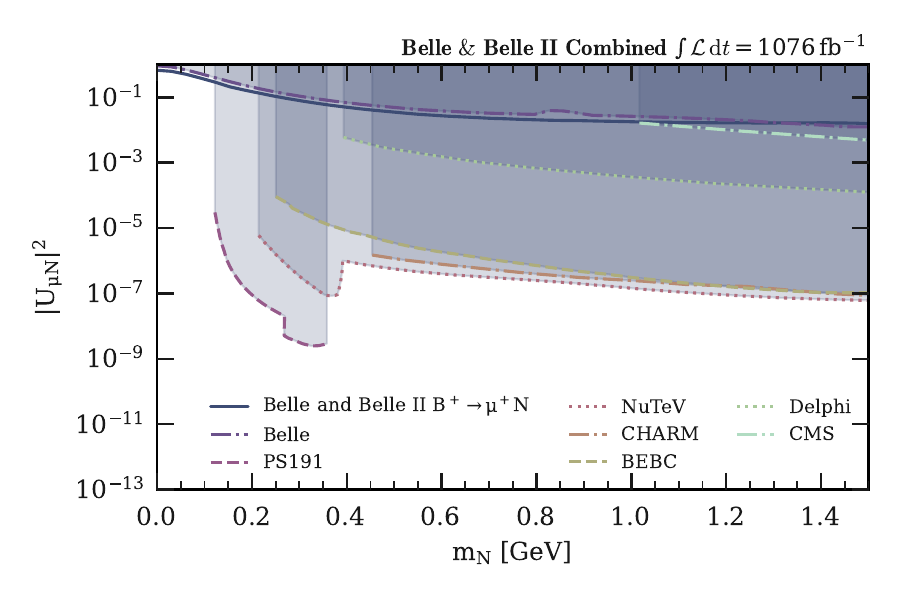}
    \caption{Excluded values of the squared mixing parameter \(|U_{\mu N}|^2\) as a function of the sterile neutrino mass \(m_N\), obtained from this analysis and compared to previous searches of sterile neutrinos. The shaded regions above the curves indicate the excluded parameter space at 90\% confidence level.}
    \label{fig:Sterile_U_Comp}
\end{figure}

The extracted number of sterile neutrino events is converted into a branching fraction of \(B^+ \to \mu^+ N\) assuming the same selection efficiency as for $B^+ \to \mu^+ \nu_\mu$. This branching fraction is then used to estimate the Bayesian upper limit for \(\mathcal{B}(B^+ \to \mu^+ N)\), as shown in Fig.~\ref{fig:Sterile_Ulocal_p}. 
The resulting upper limit is subsequently used to exclude regions in \(|U_{\mu N}|^2\)-\(m_N\) parameter space by calculating~\cite{Prim_2020,U_mu_N_values}
\begin{align}
\frac{\mathcal{B}(B^+ \to \mu^+ N)}{\mathcal{B}(B^+ \to \mu^+ \nu_\mu)} &= 
|U_{\mu N}|^2 \frac{m_N^2 + m_\mu^2}{m_\mu^2} 
\frac{\sqrt{\lambda(r_{NB}, r_{\mu B})}}{\sqrt{\lambda(0, r_{\mu B})}} \nonumber \\
&\times \frac{1 - (r_{NB}^2 - r_{\mu B}^2)^2 / (r_{NB}^2 + r_{\mu B}^2)}{1 - r_{\mu B}^2} \, ,
\end{align}
where \(|U_{\mu N}|^2\) is the coupling between the muon and the sterile neutrino, \(\lambda(x,y) = (1-(x-y)^2)(1-(x+y)^2)\) is the Käll\'{e}n function, and \(r_{X Y} = m_X / m_Y\) are mass ratios. The combined Belle and Belle II result as well as the results from previous searches by Refs.~\cite{Prim_2020,U_mu_N_PS191,U_mu_N_NuTev,U_mu_N_Charm_A,U_mu_N_Charm_B,U_mu_N_Charm_C,U_mu_N_BEBC,U_mu_N_Delphi,U_mu_N_CMS}, are shown in Fig.~\ref{fig:Sterile_U_Comp}.

\section{$B\to X_u \ell^+ \nu_\ell$ partial branching fraction measurement}\label{ref:ulnu_partial}

The $B^+ \to \mu^+ \nu_\mu$\ decay overlaps with the kinematic endpoint of \(b \to u\) transitions. This allows for a determination of the partial branching fraction \(\Delta \mathcal{B}(B \to X_u \ell^+ \nu_\ell)\) for \(p_\ell^B > \SI{2.2}{\giga\eV}\), which can be extracted simultaneously with the $B^+ \to \mu^+ \nu_\mu$\ branching fraction. Systematic uncertainties from tracking, particle identification, \(B\bar{B}\) counting, and the control-channel efficiency are included in the fit. The selection efficiencies for \(p_\mu^B > \SI{2.2}{\giga\eV}\) are \((13.69 \pm 0.57)\%\) for Belle and \((13.88 \pm 0.58)\%\) for Belle~II, with uncertainties arising from the systematic variations outlined in Section~\ref{ref:sys_uncerts}. 

The resulting partial branching fraction for \(p_\ell^B > \SI{2.2}{\giga\eV}\) is 
\begin{equation}\label{eq:partial_Xulnu}
    \Delta\mathcal{B}(B \to X_u \ell^+ \nu_\ell) 
      = (2.72 \pm 0.05 \pm 0.29)\times 10^{-4} \, ,
\end{equation}
where the first uncertainty is statistical and the second systematic. Table~\ref{table:sys_table_comb_Xu} in Appendix~\ref{sec:appendix_A} summarizes the systematic uncertainties affecting the measurement. This result agrees within two standard deviations with the measurement in Ref.~\cite{Babar_partial}, \(\Delta\mathcal{B}(B \to X_u \ell^+ \nu_\ell) = (0.33 \pm 0.04) \times 10^{-3}\), but achieves improved precision. Extrapolating to the full phase space using the weighted mean of the Belle and Belle II efficiencies yields 
\begin{equation} 
    \mathcal{B}(B \to X_u \ell^+ \nu_\ell) = (1.97 \pm 0.07 \pm 0.22) \times 10^{-3} \, ,  
\end{equation}  
which is consistent and has similar precision to the HFLAV average~\cite{HFLAV_2024}, \(\mathcal{B}(B \to X_u \ell^+ \nu_\ell) = (1.92 \pm 0.21) \times 10^{-3}\).

\section{Investigation of weak annihilation process}\label{ref:wa_proc}

\begin{figure}[htbp]
    \centering
    \includegraphics[width=\linewidth]{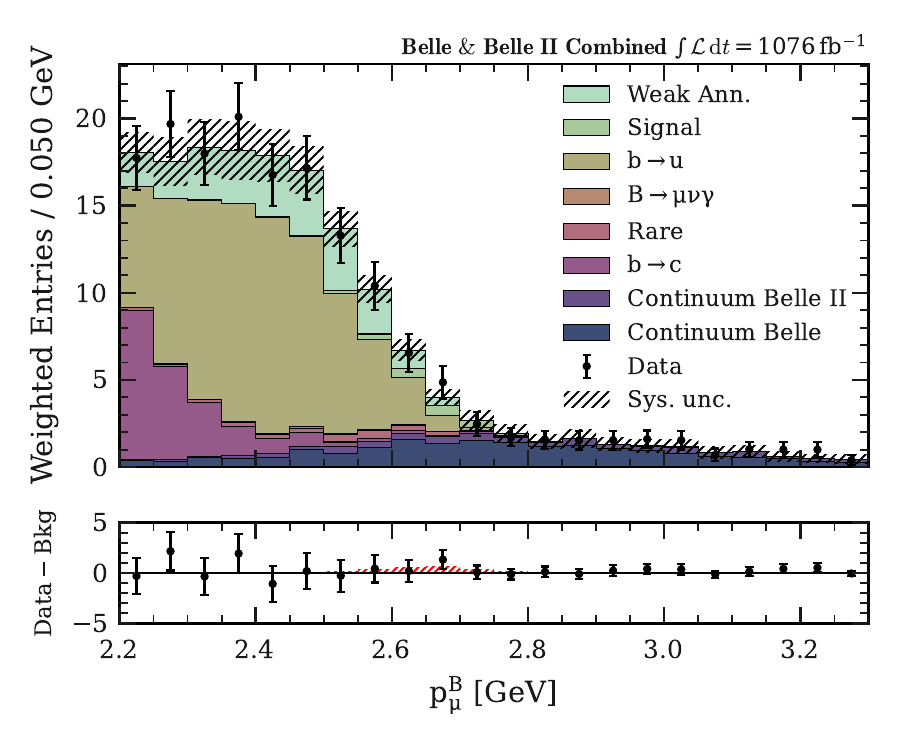}
    \caption{Weighted distribution of the fitted muon momentum in the \(B\) rest frame, with an additional template for weak annihilation events included using the same weights derived for Fig.~\ref{fig:weighted_fit}.}
    \label{fig:wa_fit_eight_cats_weighted}
\end{figure}

A model-dependent search for the weak-annihilation (WA) process is performed using the fitting strategy and templates described in Section~\ref{Fit_setup}. For this, the yields \(B^+ \to \mu^+ \nu_\mu\), \(B \to X_u \ell^+ \nu_\ell\), and weak annihilation, as well as the parameter \(\alpha\) are unconstrained. Figure~\ref{fig:wa_fit_eight_cats_weighted} shows the combined muon-momentum distribution, obtained by weighting the signal-enriched categories to enhance sensitivity to a potential WA contribution. The corresponding fit results for all eight classifier categories are presented in Appendix~\ref{sec:appendix_A} in Figure~\ref{fig:wa_fit_eight_cats}.

The extracted WA branching fraction is determined to be
\begin{align}
    \mathcal{B}(B^+ \overset{\text{WA}}{\to} X_u \ell^+  \nu_\ell) = (5.76 \pm 3.92) \times 10^{-5}  \, ,
\end{align}
where the error is the sum of statistical and systematic uncertainties. This translates to \(\nu^{\text{WA}} = 3600 \pm 2473\) WA events, which can substitute a large fraction of \(B \to X_u \ell^+ \nu_\ell\) events, leading to a strong anti-correlation between the two processes. Including WA in the fit alters the extracted partial and full branching fractions: 
\begin{align}
    \Delta\mathcal{B}(B \to X_u \ell^+ \nu_\ell) &= (2.06 \pm 0.44) \times 10^{-4} \, , \\
    \mathcal{B}(B^+ \to \mu^+ \nu_\mu) &= (3.4 \pm 2.2) \times 10^{-7} \, , 
\end{align}
with the uncertainties indicating the sum of statistical and systematic uncertainties. 

\begin{figure}[htbp]
    \centering
    \includegraphics[width=\linewidth]{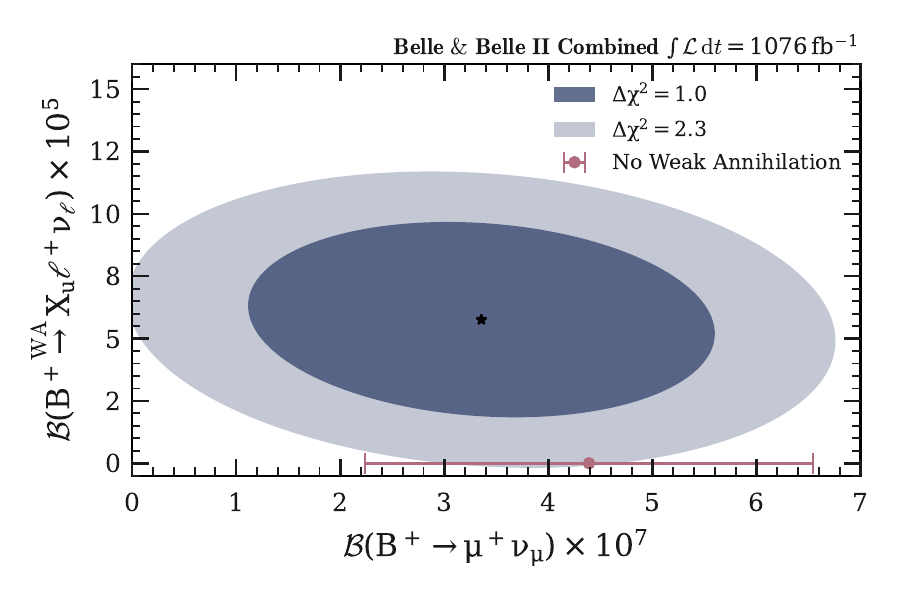}
    \includegraphics[width=\linewidth]{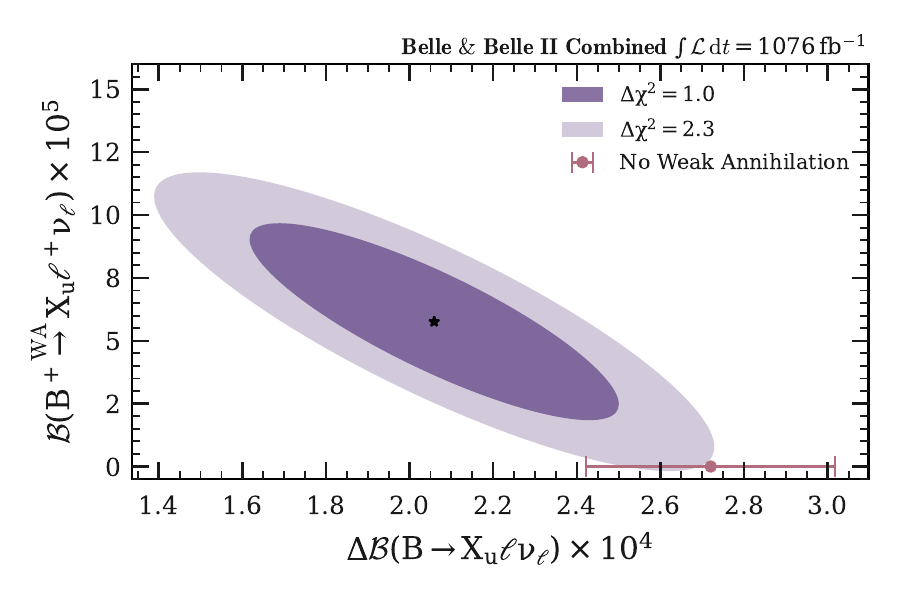}
    \caption{Two-dimensional likelihood contours showing the correlation between the WA contribution and the extracted $B^+ \to \mu^+ \nu_\mu$ branching fraction (top), and the partial branching fraction for \( B \to X_u \ell^+ \nu_\ell \) (bottom). The contours correspond to regions of constant \(\Delta\chi^2 = 1.0\) and \(\Delta\chi^2 = 2.3\), approximately representing 38.3\% and 68.3\% confidence levels. The rose points with error bars correspond to the measured $B^+ \to \mu^+ \nu_\mu$ and partial \(B \to X_u \ell^+ \nu_\ell\) branching fraction, as reported in Eq.~\ref{eq:final_bf} and Eq.~\ref{eq:partial_Xulnu} respectively, if there is no WA contribution included.}
    \label{fig:WA_Munu}
\end{figure}

\begin{figure}[htbp]
    \centering
    \includegraphics[width=\linewidth]{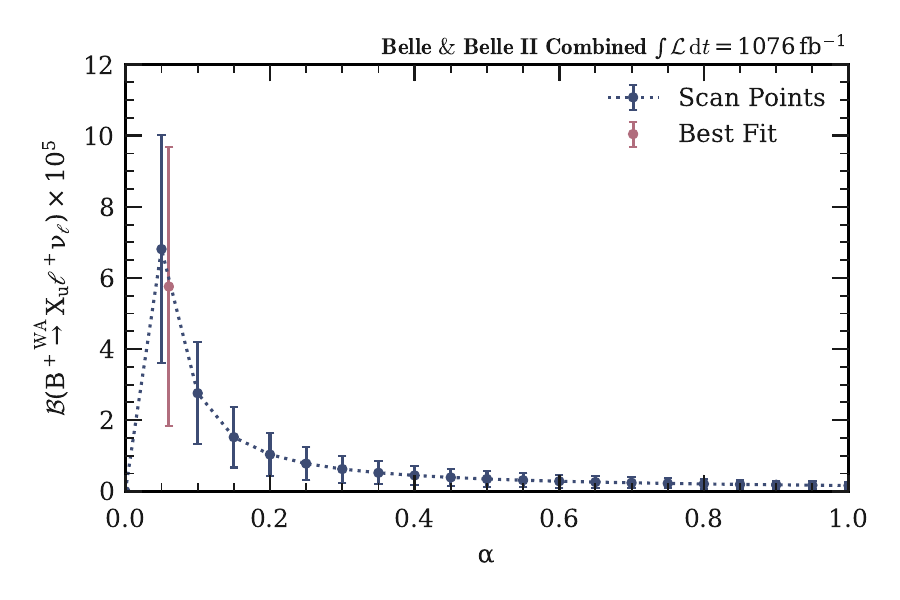}
\caption{
Weak annihilation branching fraction as a function of the parameter \(\alpha\), which determines the shape of the  muon momentum distribution in the \(B\) rest frame. For each value of \(\alpha\), the fit is performed with the parameter fixed at that value extracting the number of WA events. The black points show the scan results, while the rose point indicates the best fit value.}
    \label{fig:Morph_par}
\end{figure}

Figure~\ref{fig:WA_Munu} shows the two-dimensional likelihood contours in the parameter space of $\mathcal{B}(B^+ \overset{\text{WA}}{\to} X_u \ell^+ \nu_\ell)$ versus the branching fractions of $B^+ \to \mu^+ \nu_\mu$ and $B \to X_u \ell^+ \nu_\ell$. When allowing the model parameter $\alpha$ to float, a one-sided significance of 2.4 standard deviations relative to the background-only hypothesis is obtained. If the parameter is instead fixed to scan across different model assumptions, a substantially smaller WA branching fraction is found in most models. The corresponding branching fractions are summarized in Figure~\ref{fig:Morph_par}. We refrain from drawing stronger conclusions due to the strong anti-correlation with $B \to X_u \ell^+ \nu_\ell$ decays, as the WA contribution becomes significant only in scenarios with a flat $p_\mu^B$ distribution that closely resembles the kinematics of $B \to X_u \ell^+ \nu_\ell$ decays. To gain further insight, a more precise theoretical understanding of the kinematic shape of the weak-annihilation contribution is required.

\section{Summary and conclusion}\label{ref:conclusion}

In this paper, we report the result of a search for $B^+ \to \mu^+ \nu_\mu$\ decays using \SI{1076}{\per\femto\barn} of data collected with the Belle and Belle~II detectors at the KEKB and SuperKEKB \(e^+e^-\) colliders at a center-of-mass energy of \SI{10.58}{\giga\eV}. This search includes a reanalysis of the Belle data, which supersedes the previous measurement~\cite{Prim_2020}, and provides the first Belle~II result on the $B^+ \to \mu^+ \nu_\mu$\ branching fraction. An inclusive tagging approach is employed to reconstruct and calibrate the kinematics of the tag-side \(B\) meson, enabling measurement of the muon momentum \(p_\mu^B\) in the signal-side \(B\) meson rest frame with improved resolution relative to the center-of-mass frame. The $B^+ \to \mu^+ \nu_\mu$\ branching fraction is extracted from a binned maximum-likelihood fit to \(p_\mu^B\) and no contributions from weak-annihilation processes are included. The result is 
\begin{align}
    \mathcal{B}(B^+ \to \mu^+ \nu_\mu) = (4.4 \pm 1.9 \pm 1.0) \times 10^{-7} \, ,
\end{align}
with a significance of 2.4 standard deviations relative to the background-only hypothesis and in agreement with the expected sensitivity. The systematic uncertainties due to the finite size of the MC samples, including those from the limited size of the generated sample used to model the continuum contribution, will decrease with the simulation of larger samples. Auxiliary measurements using additional data can further constrain the input parameters of the form factor models, branching fractions, and the inclusive model describing the \(b \to u\) contribution, thereby reducing the associated systematic uncertainties. Additional data will also reduce the uncertainties associated with particle identification and tracking efficiencies, as well as the total number of \(B \bar B\) pairs.

As no significant signal is observed, upper limits at 90\% credibility and confidence level, respectively, are set using Bayesian and frequentist methods: 
\begin{align}
    \mathcal{B}(B^+ \to \mu^+ \nu_\mu) &< 7.2 \times 10^{-7}, \\
    \mathcal{B}(B^+ \to \mu^+ \nu_\mu) &< 6.7 \times 10^{-7},
\end{align}
respectively. The result constitutes the most sensitive search for the decay \(B^+ \to \mu^+ \nu_\mu\) to date.

The measured branching fraction is interpreted to constrain the parameter space of type~II and type~III two-Higgs-doublet models, excluding significant regions of their parameter space. A search is carried out for massive sterile neutrinos in the range \mbox{\SI{0.0}{\giga\eV} \(\leq m_N \leq\) \SI{1.5}{\giga\eV}}. No significant excess is observed, with the largest deviation from the background-only hypothesis being 0.6 standard deviations, taking into account the look-elsewhere effect. The observed yields are translated into limits on the squared mixing parameter \(|U_{\mu N}|^2\) as a function of the sterile neutrino mass \(m_N\) and cover a wider parameter space. 

The partial branching fraction of semileptonic $B \to X_u \ell^+ \nu_\ell$ decays with \(p_\mu^B > \) \SI{2.2}{\giga\eV} is measured to be
\begin{equation}
    \Delta\mathcal{B}(B \to X_u \ell^+ \nu_\ell) = (2.72 \pm 0.05 \pm 0.29) \times 10^{-4} \, ,
\end{equation}
where uncertainties from the theoretical model and selection efficiency are included. When extrapolated to the full phase-space, this result and its experimental uncertainty are consistent with, and of comparable precision to, the world average reported in Ref.~\cite{HFLAV_2024}.

A model-dependent search for weak-annihilation decays is carried out. Because the branching fraction and spectral shape of such decays are poorly known, the WA component is modeled as a combination of flat and peaking contributions with a single floating fraction. We find
\begin{align}
    \mathcal{B}(B^+ \overset{\text{WA}}{\to} X_u \ell^+  \nu_\ell) = (5.76 \pm 3.92) \times 10^{-5}  \, ,
\end{align}
with a significance of 2.4 standard deviations above the background-only hypothesis. The resulting WA template closely resembles the kinematic shape of $B \to X_u \ell \nu_\ell$ decays, leading to a strong anti-correlation between the two contributions. When alternative values of the model fraction are chosen, the fitted WA branching fractions become much smaller and less significant. Including the WA template, reduces the observed $B^+ \to \mu^+ \nu_\mu$ branching fraction by about 23\%. To gain further insights, a more precise theoretical understanding of the kinematic shape of the weak-annihilation contribution is required. 

\section*{Acknowledgements}
This work, based on data collected using the Belle II detector, which was built and commissioned prior to March 2019,
and data collected using the Belle detector, which was operated until June 2010,
was supported by
Higher Education and Science Committee of the Republic of Armenia Grant No.~23LCG-1C011;
Australian Research Council and Research Grants
No.~DP200101792, 
No.~DP210101900, 
No.~DP210102831, 
No.~DE220100462, 
No.~LE210100098, 
and
No.~LE230100085; 
Austrian Federal Ministry of Education, Science and Research,
Austrian Science Fund (FWF) Grants
DOI:~10.55776/P34529,
DOI:~10.55776/J4731,
DOI:~10.55776/J4625,
DOI:~10.55776/M3153,
and
DOI:~10.55776/PAT1836324,
and
Horizon 2020 ERC Starting Grant No.~947006 ``InterLeptons'';
Natural Sciences and Engineering Research Council of Canada, Digital Research Alliance of Canada, and Canada Foundation for Innovation;
National Key R\&D Program of China under Contract No.~2024YFA1610503,
and
No.~2024YFA1610504
National Natural Science Foundation of China and Research Grants
No.~11575017,
No.~11761141009,
No.~11705209,
No.~11975076,
No.~12135005,
No.~12150004,
No.~12161141008,
No.~12405099,
No.~12475093,
and
No.~12175041,
and Shandong Provincial Natural Science Foundation Project~ZR2022JQ02;
the Czech Science Foundation Grant No. 22-18469S,  Regional funds of EU/MEYS: OPJAK
FORTE CZ.02.01.01/00/22\_008/0004632 
and
Charles University Grant Agency project No. 246122;
European Research Council, Seventh Framework PIEF-GA-2013-622527,
Horizon 2020 ERC-Advanced Grants No.~267104 and No.~884719,
Horizon 2020 ERC-Consolidator Grant No.~819127,
Horizon 2020 Marie Sklodowska-Curie Grant Agreement No.~700525 ``NIOBE''
and
No.~101026516,
and
Horizon 2020 Marie Sklodowska-Curie RISE project JENNIFER2 Grant Agreement No.~822070 (European grants);
L'Institut National de Physique Nucl\'{e}aire et de Physique des Particules (IN2P3) du CNRS
and
L'Agence Nationale de la Recherche (ANR) under Grant No.~ANR-23-CE31-0018 (France);
BMFTR, DFG, HGF, MPG, and AvH Foundation (Germany);
Department of Atomic Energy under Project Identification No.~RTI 4002,
Department of Science and Technology,
and
UPES SEED funding programs
No.~UPES/R\&D-SEED-INFRA/17052023/01 and
No.~UPES/R\&D-SOE/20062022/06 (India);
Israel Science Foundation Grant No.~2476/17,
U.S.-Israel Binational Science Foundation Grant No.~2016113, and
Israel Ministry of Science Grant No.~3-16543;
Istituto Nazionale di Fisica Nucleare and the Research Grants BELLE2,
and
the ICSC – Centro Nazionale di Ricerca in High Performance Computing, Big Data and Quantum Computing, funded by European Union – NextGenerationEU;
Japan Society for the Promotion of Science, Grant-in-Aid for Scientific Research Grants
No.~16H03968,
No.~16H03993,
No.~16H06492,
No.~16K05323,
No.~17H01133,
No.~17H05405,
No.~18K03621,
No.~18H03710,
No.~18H05226,
No.~19H00682, 
No.~20H05850,
No.~20H05858,
No.~22H00144,
No.~22K14056,
No.~22K21347,
No.~23H05433,
No.~26220706,
and
No.~26400255,
and
the Ministry of Education, Culture, Sports, Science, and Technology (MEXT) of Japan;  
National Research Foundation (NRF) of Korea Grants 
No.~2021R1-A6A1A-03043957,
No.~2021R1-F1A-1064008, 
No.~2022R1-A2C-1003993,
No.~2022R1-A2C-1092335,
No.~RS-2016-NR017151,
No.~RS-2018-NR031074,
No.~RS-2021-NR060129,
No.~RS-2023-00208693,
No.~RS-2024-00354342
and
No.~RS-2025-02219521,
Radiation Science Research Institute,
Foreign Large-Size Research Facility Application Supporting project,
the Global Science Experimental Data Hub Center, the Korea Institute of Science and
Technology Information (K25L2M2C3 ) 
and
KREONET/GLORIAD;
Universiti Malaya RU grant, Akademi Sains Malaysia, and Ministry of Education Malaysia;
Frontiers of Science Program Contracts
No.~FOINS-296,
No.~CB-221329,
No.~CB-236394,
No.~CB-254409,
and
No.~CB-180023, and SEP-CINVESTAV Research Grant No.~237 (Mexico);
the Polish Ministry of Science and Higher Education and the National Science Center;
the Ministry of Science and Higher Education of the Russian Federation
and
the HSE University Basic Research Program, Moscow;
University of Tabuk Research Grants
No.~S-0256-1438 and No.~S-0280-1439 (Saudi Arabia), and
Researchers Supporting Project number (RSPD2025R873), King Saud University, Riyadh,
Saudi Arabia;
Slovenian Research Agency and Research Grants
No.~J1-50010
and
No.~P1-0135;
Ikerbasque, Basque Foundation for Science,
State Agency for Research of the Spanish Ministry of Science and Innovation through Grant No. PID2022-136510NB-C33, Spain,
Agencia Estatal de Investigacion, Spain
Grant No.~RYC2020-029875-I
and
Generalitat Valenciana, Spain
Grant No.~CIDEGENT/2018/020;
the Swiss National Science Foundation;
The Knut and Alice Wallenberg Foundation (Sweden), Contracts No.~2021.0174, No.~2021.0299, and No.~2023.0315;
National Science and Technology Council,
and
Ministry of Education (Taiwan);
Thailand Center of Excellence in Physics;
TUBITAK ULAKBIM (Turkey);
National Research Foundation of Ukraine, Project No.~2020.02/0257,
and
Ministry of Education and Science of Ukraine;
the U.S. National Science Foundation and Research Grants
No.~PHY-1913789 
and
No.~PHY-2111604, 
and the U.S. Department of Energy and Research Awards
No.~DE-AC06-76RLO1830, 
No.~DE-SC0007983, 
No.~DE-SC0009824, 
No.~DE-SC0009973, 
No.~DE-SC0010007, 
No.~DE-SC0010073, 
No.~DE-SC0010118, 
No.~DE-SC0010504, 
No.~DE-SC0011784, 
No.~DE-SC0012704, 
No.~DE-SC0019230, 
No.~DE-SC0021616, 
No.~DE-SC0022350, 
No.~DE-SC0023470; 
and
the Vietnam Academy of Science and Technology (VAST) under Grants
No.~NVCC.05.02/25-25
and
No.~DL0000.05/26-27.

These acknowledgements are not to be interpreted as an endorsement of any statement made
by any of our institutes, funding agencies, governments, or their representatives.

We thank the SuperKEKB team for delivering high-luminosity collisions;
the KEK cryogenics group for the efficient operation of the detector solenoid magnet and IBBelle on site;
the KEK Computer Research Center for on-site computing support; the NII for SINET6 network support;
and the raw-data centers hosted by BNL, DESY, GridKa, IN2P3, INFN, 
PNNL/EMSL, 
and the University of Victoria.

\bibliography{references}

\onecolumngrid
\appendix

\section{Continuum-Suppression BDT Input Variables}\label{sec:appendix_A_0}
The continuum-suppression BDT input variables are listed below in descending order of their discriminating power.
\begin{itemize}
    \item Cosine of the angle between the signal muon momentum and thrust axis of the ROE.
    \item 4th order harmonic moment with respect to the thrust axis.
    \item 4th order normalized modified Fox--Wolfram moment of type \textit{so} for missing momentum. When particle \(i\)
and particle \(j\) both are from the ROE type \textit{oo} is used, while type \textit{so} is used if particle \(i\) is related to the signal-\(B\) meson.~\cite{Belle:2003fgr}
    \item Magnitude of the thrust axis of the ROE.
    \item Normalized beam-constrained mass \(\hat{m}_{\mathrm{bc}}^{\text{tag}}\).
    \item 2nd order normalized modified Fox--Wolfram moment of type \textit{so} for neutral particles.
    \item Cosine of the polar angle of the missing momentum calculated in the lab frame
    \item Normalized energy difference \(\Delta \hat{E}\)
    \item 2nd order normalized modified Fox--Wolfram moment of type \textit{so} for charged particles.
    \item Transverse energy.
    \item Cosine of the polar angle of the thrust axis.
    \item 1st order harmonic moment with respect to the thrust axis.
    \item 2nd CLEO cone.~\cite{Cleo_Cones}
    \item 4th order normalized modified Fox--Wolfram moment of type \textit{so} for neutral particles.
    \item 0th order normalized modified Fox--Wolfram moment of type \textit{so} for neutral particles.
    \item 4th order normalized modified Fox--Wolfram moment of type \textit{oo}
\end{itemize}

\clearpage

\section{Continuum Correction using Off-Resonance Data}\label{sec:appendix_A_1}

\begin{figure}[htbp]
    \centering
    \includegraphics[width=0.48\linewidth]{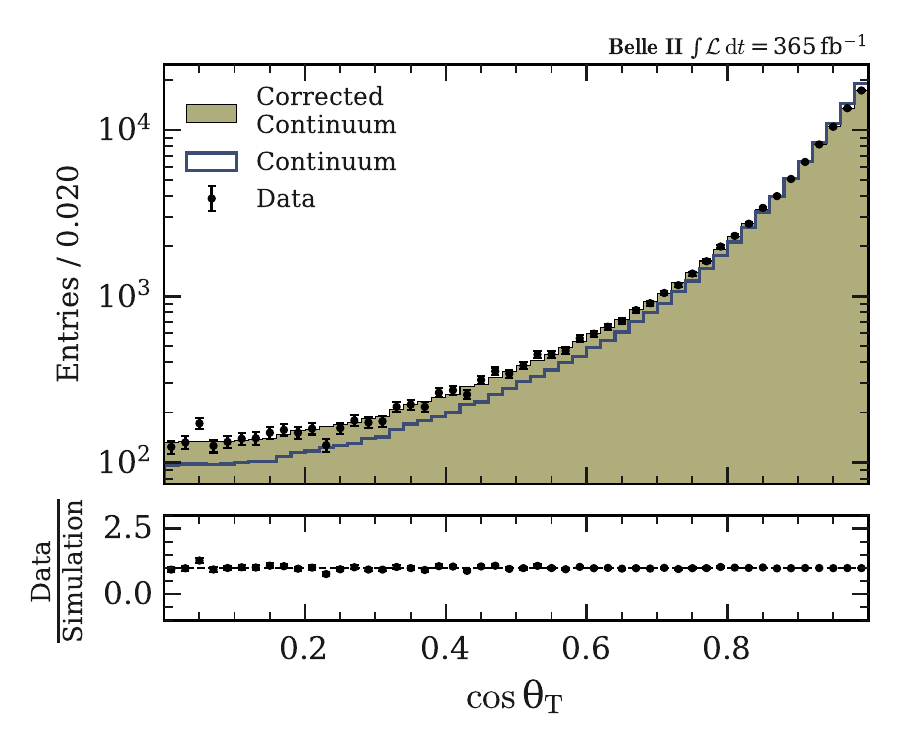}
    \includegraphics[width=0.48\linewidth]{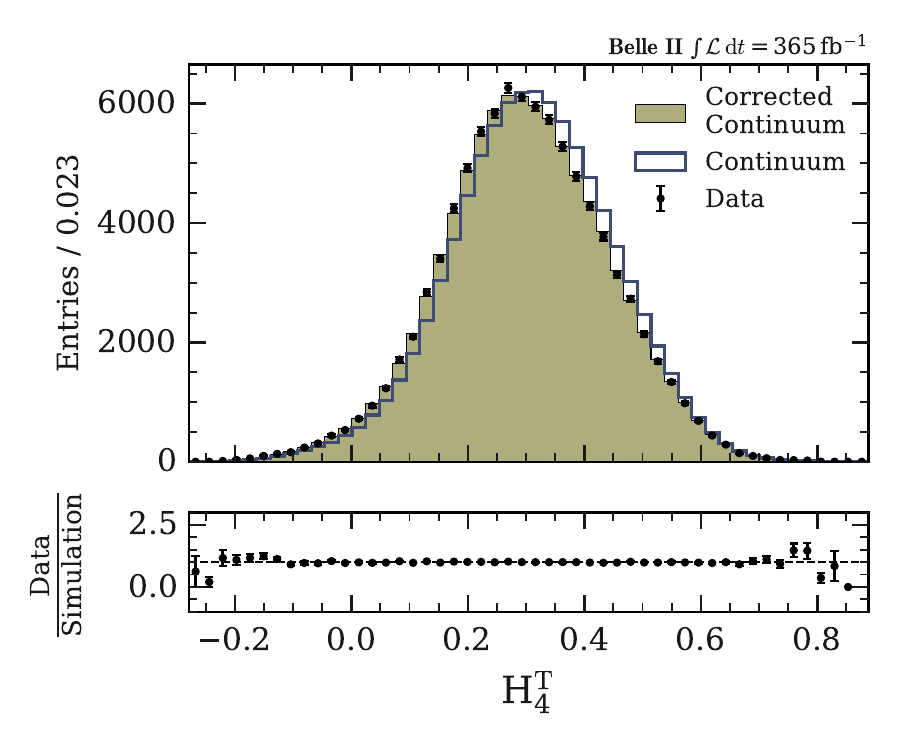}
    \includegraphics[width=0.48\linewidth]{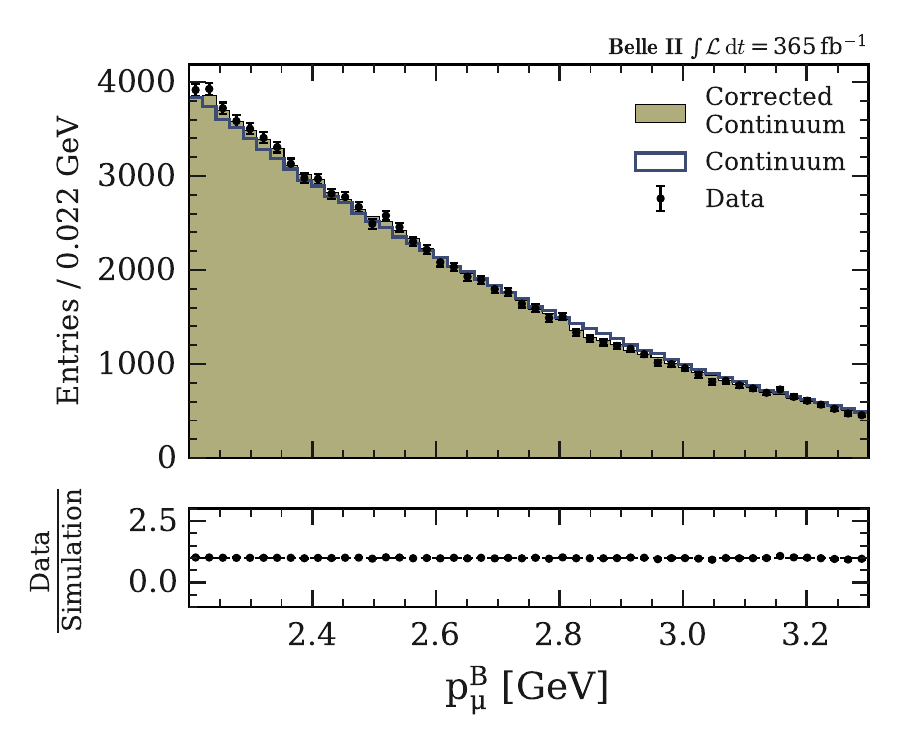}
    \includegraphics[width=0.48\linewidth]{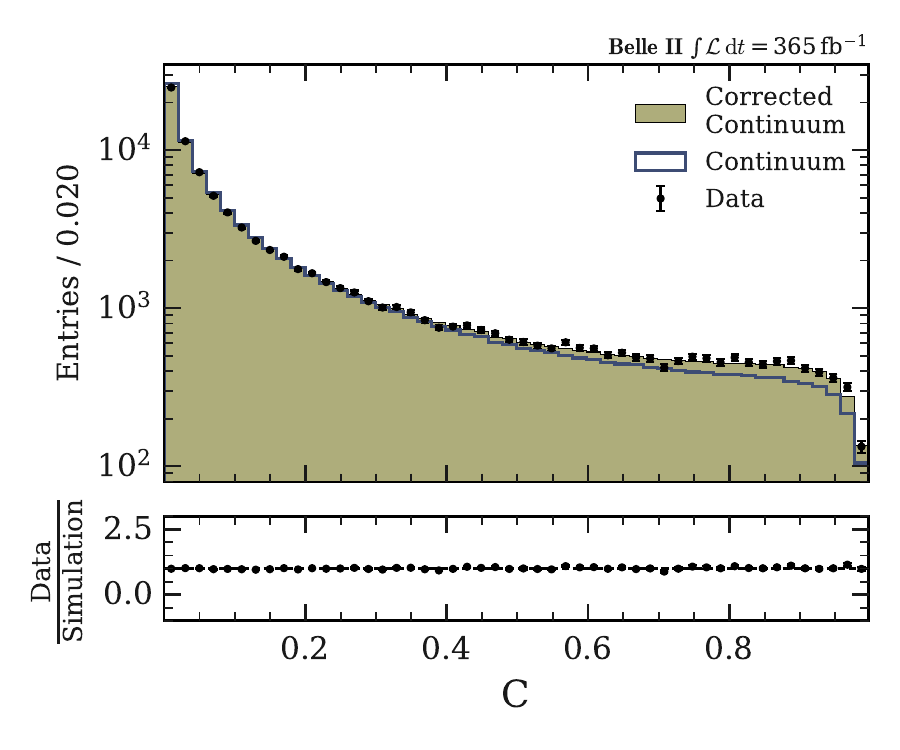}
    \caption{Corrected continuum distributions of the two leading BDT input variables, the muon momentum in the signal-\(B\) meson rest frame frame, and the BDT output classifier, compared to off-resonance data. The uncorrected continuum distribution is shown in dark blue.}
    \label{fig:appendix_A_1}
\end{figure}

\section{Fitted Distributions}\label{sec:appendix_A}

\begin{figure}[htbp]
    \centering
    \includegraphics[height=0.23\textheight]{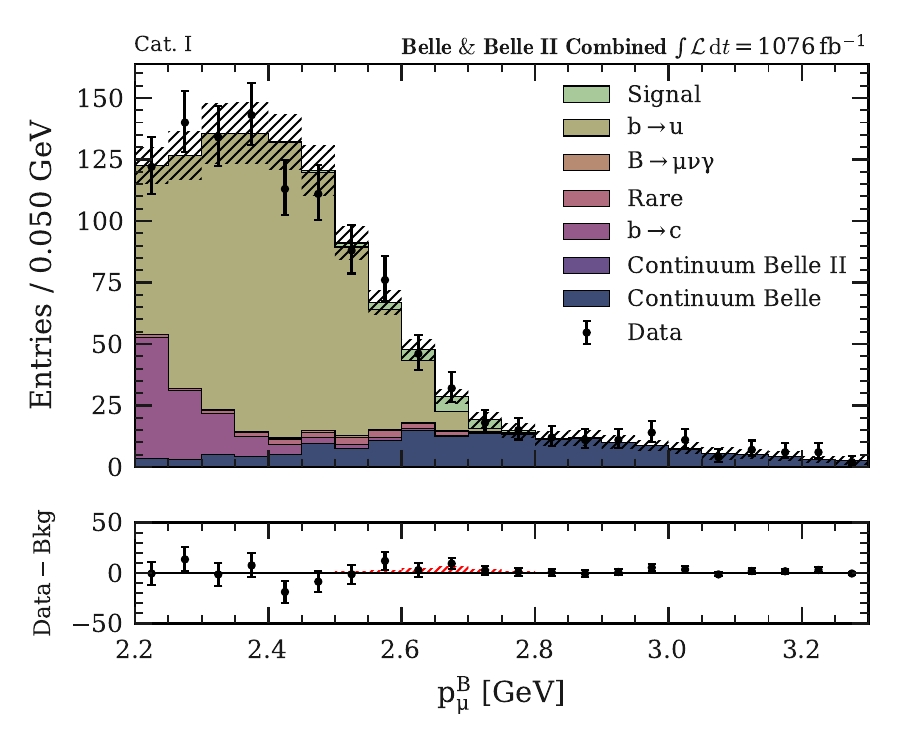}
    \includegraphics[height=0.23\textheight]{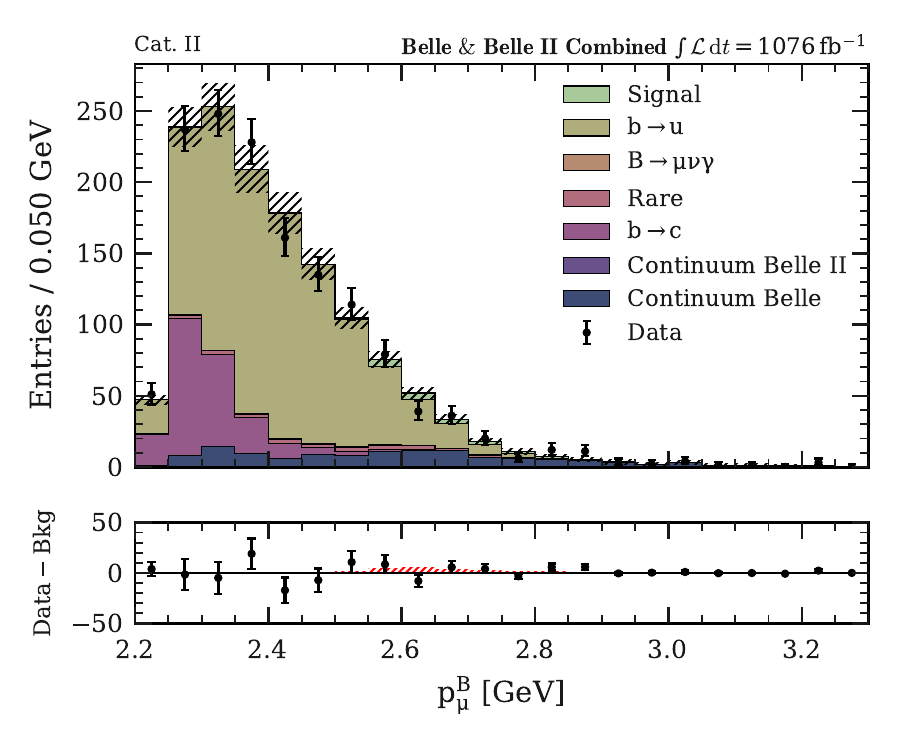}
    \includegraphics[height=0.23\textheight]{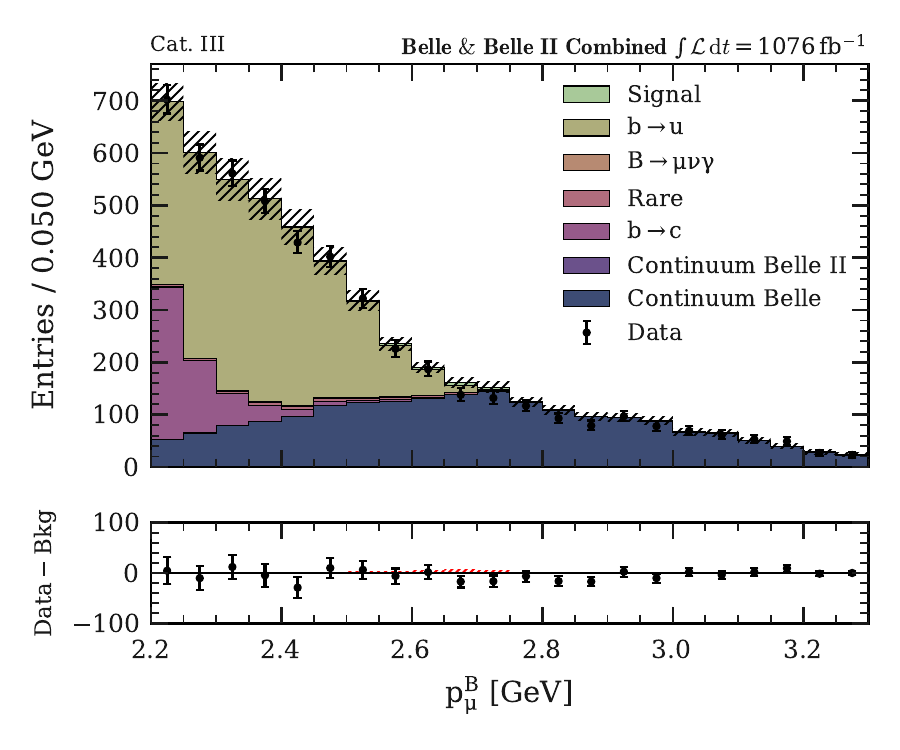}
    \includegraphics[height=0.23\textheight]{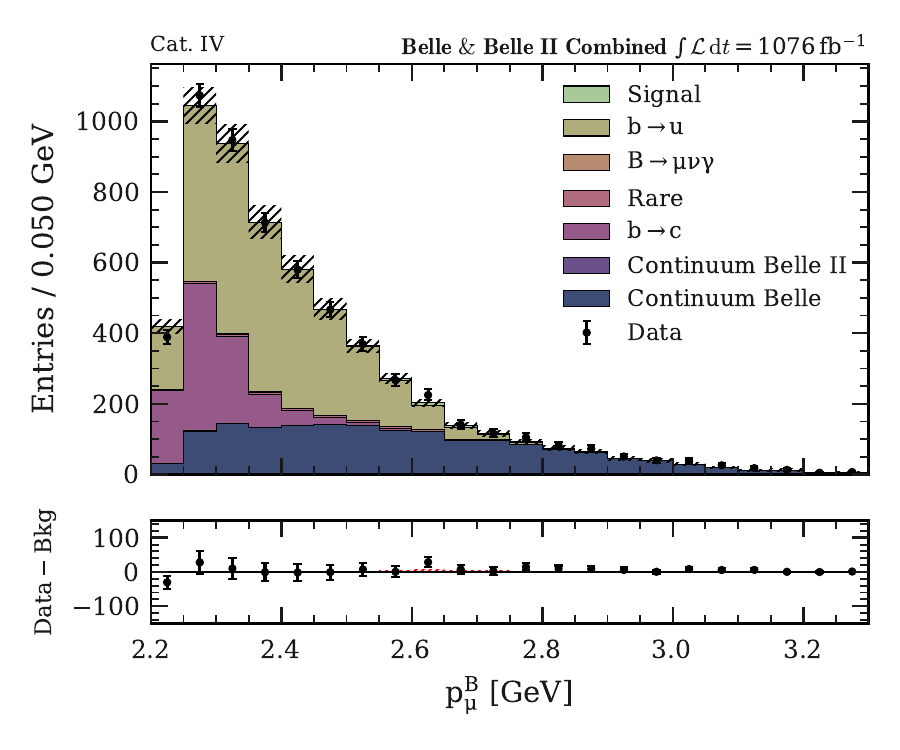}
    \includegraphics[height=0.23\textheight]{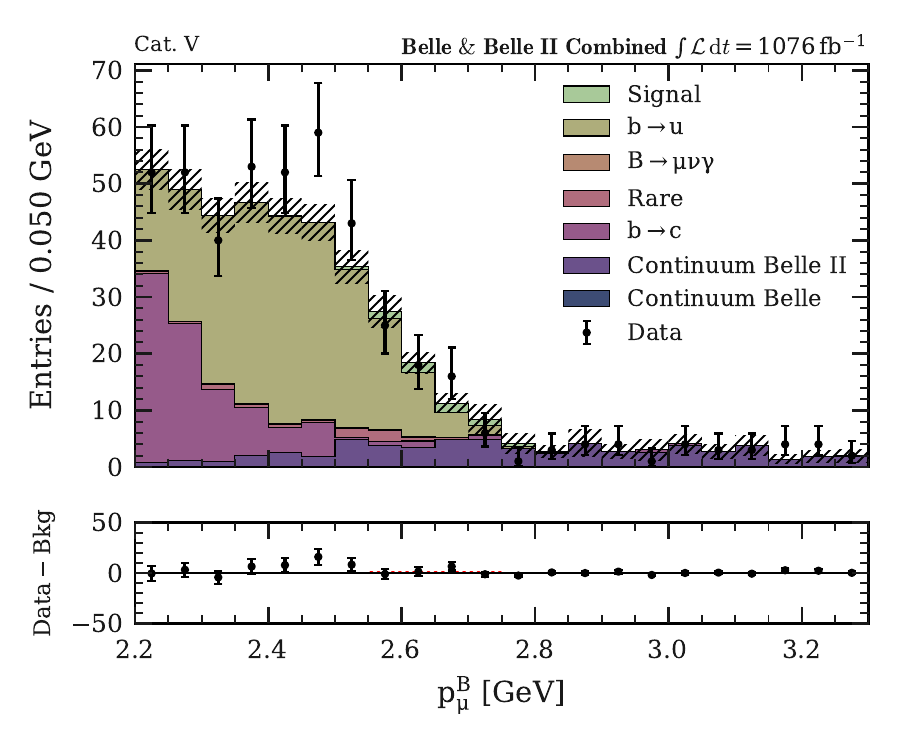}
    \includegraphics[height=0.23\textheight]{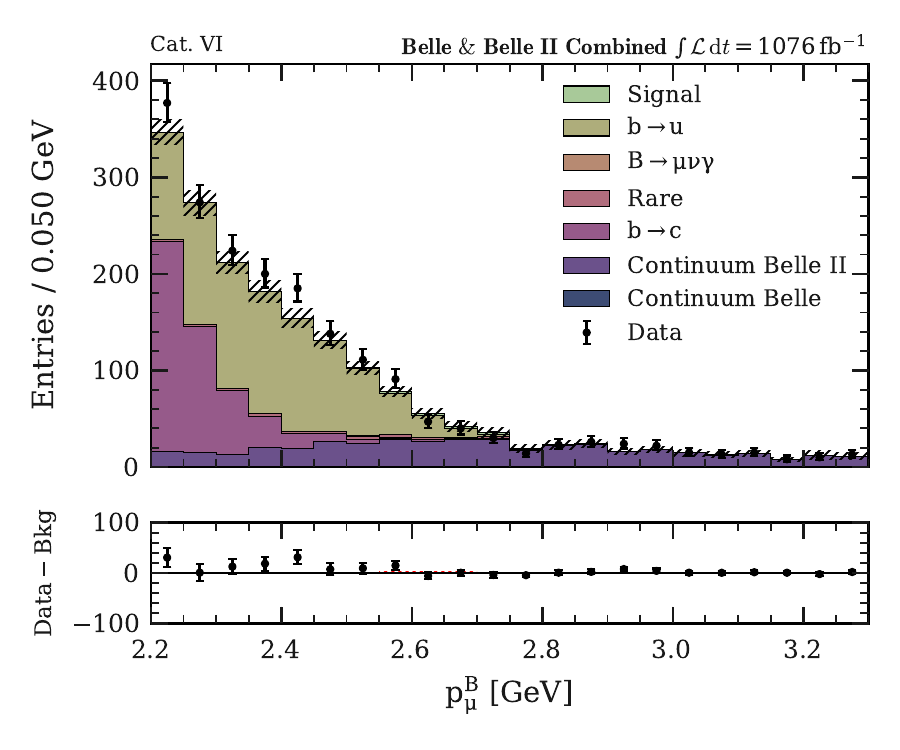}
    \includegraphics[height=0.23\textheight]{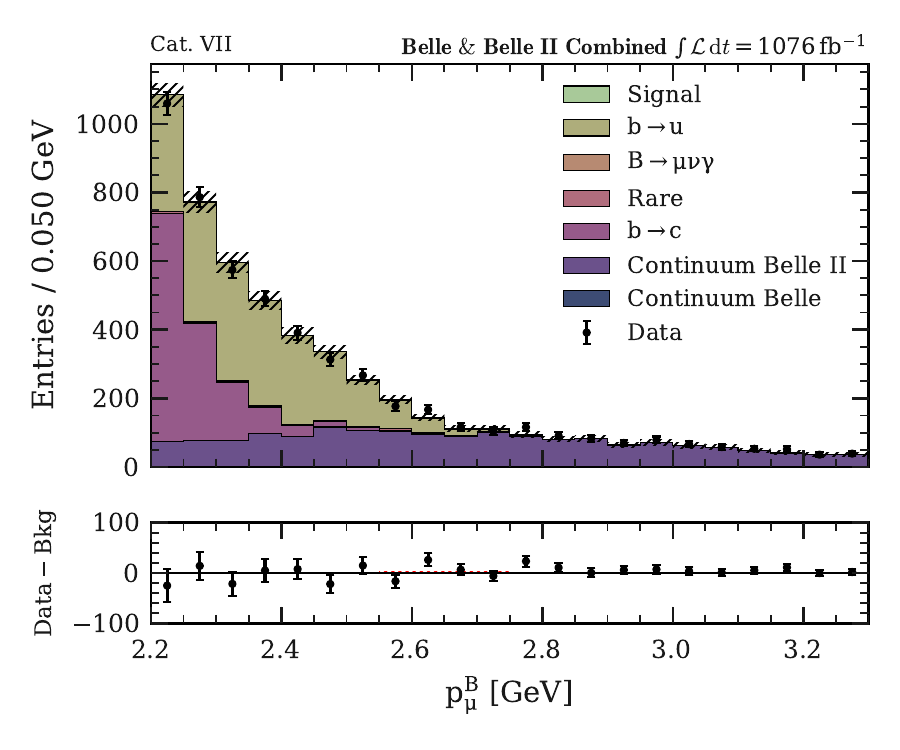}
    \includegraphics[height=0.23\textheight]{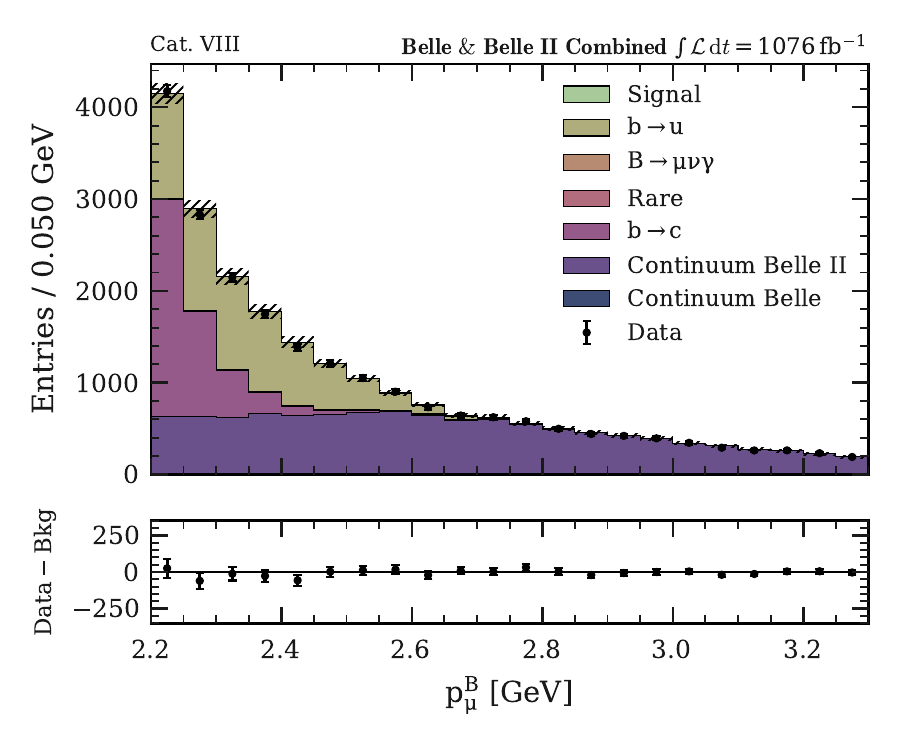}
    \caption{Distribution of the fitted muon momentum in the \(B\) rest frame for the eight categories. The stacked histograms shows the fitted signal and background processes, the data points the combined Belle and Belle II sample and the hatched band the systematic uncertainty.}
    \label{fig:fit_eight_cats}
\end{figure}

\begin{figure}[htbp]
    \centering
    \includegraphics[height=0.23\textheight]{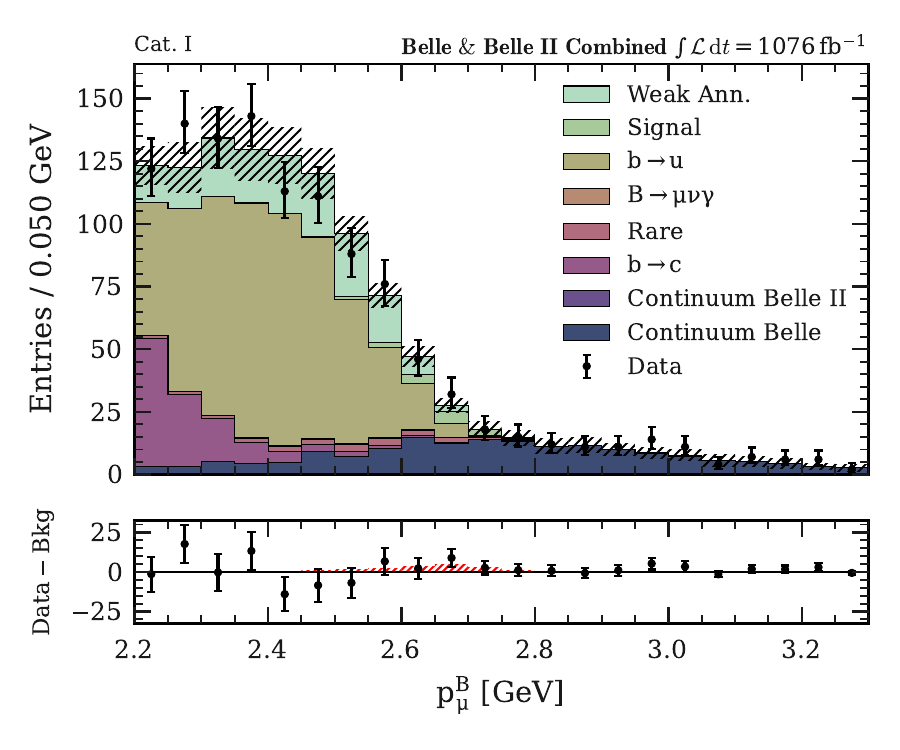}
    \includegraphics[height=0.23\textheight]{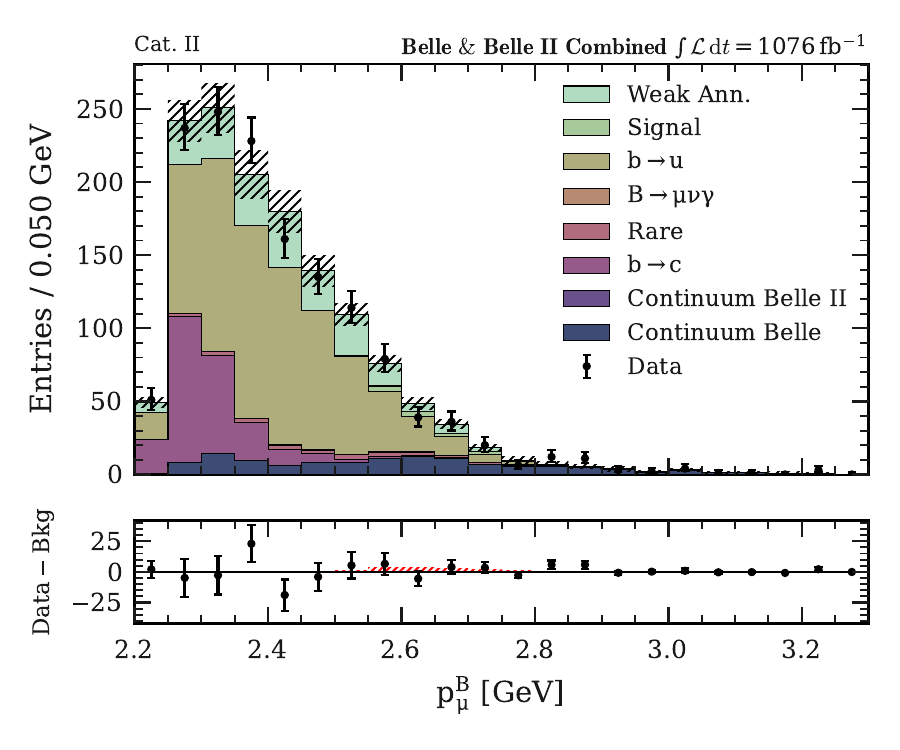}
    \includegraphics[height=0.23\textheight]{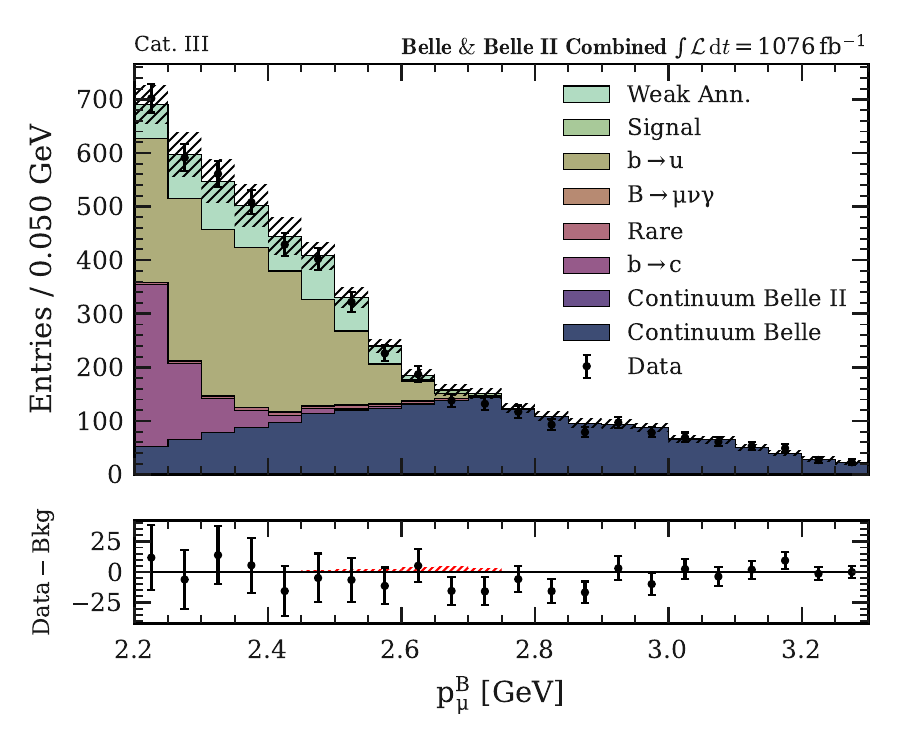}
    \includegraphics[height=0.23\textheight]{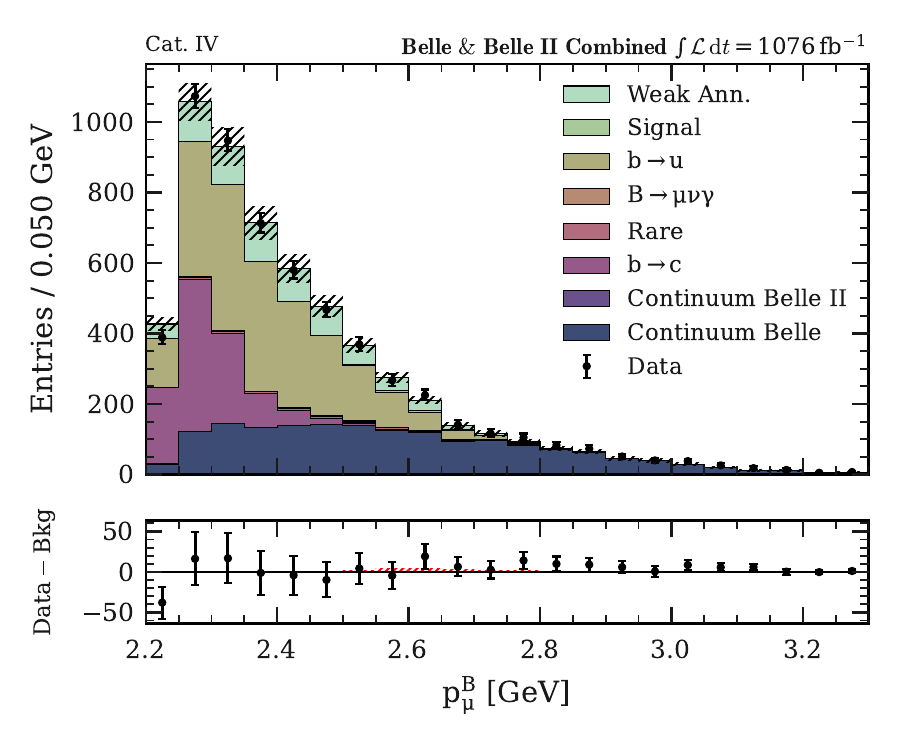}
    \includegraphics[height=0.23\textheight]{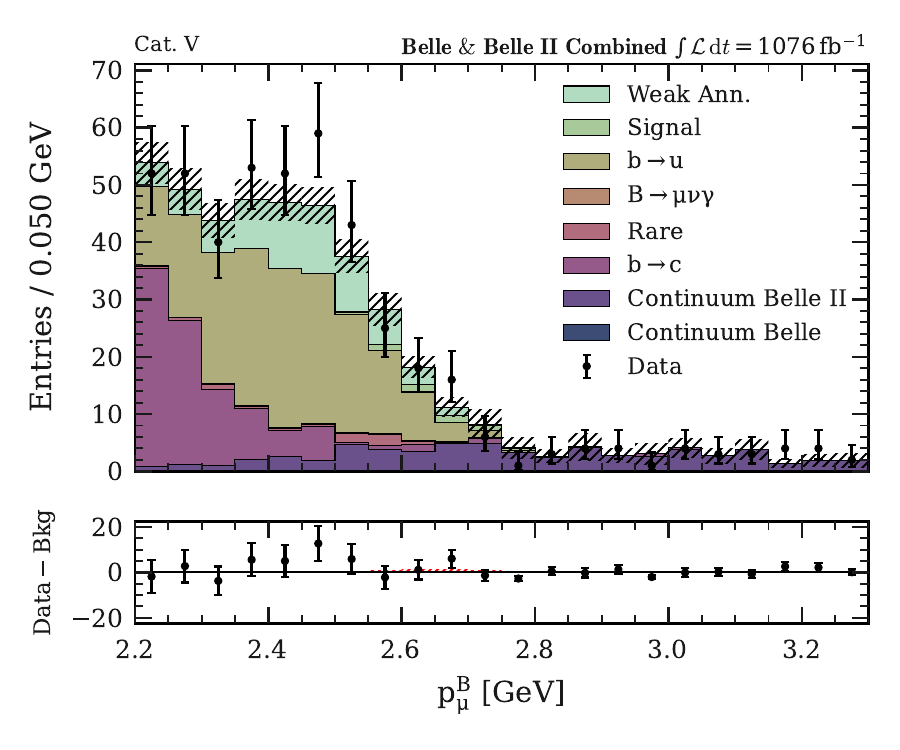}
    \includegraphics[height=0.23\textheight]{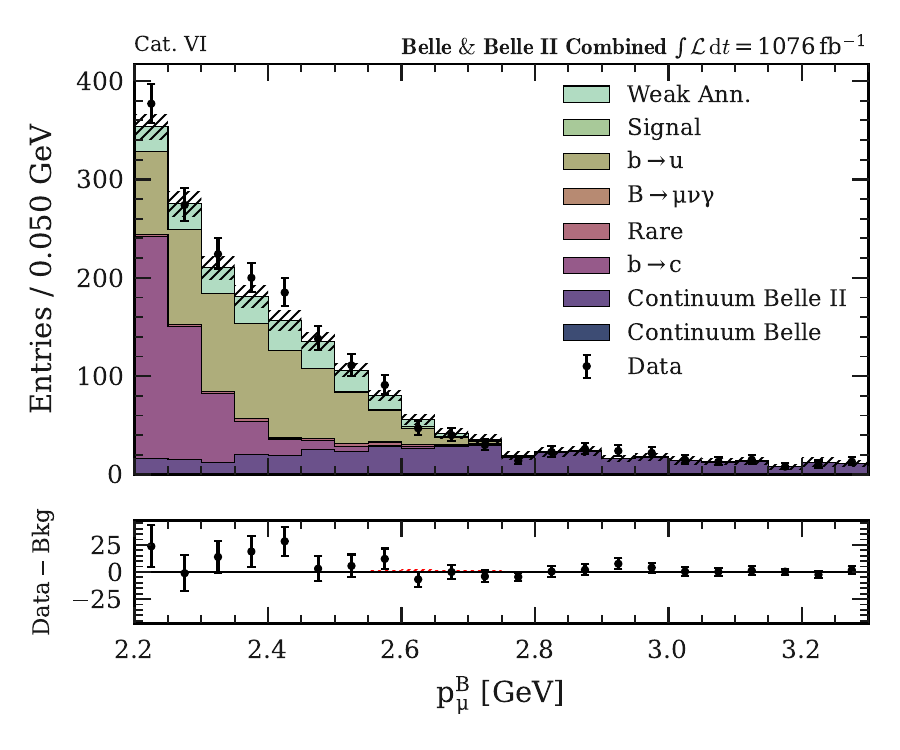}
    \includegraphics[height=0.23\textheight]{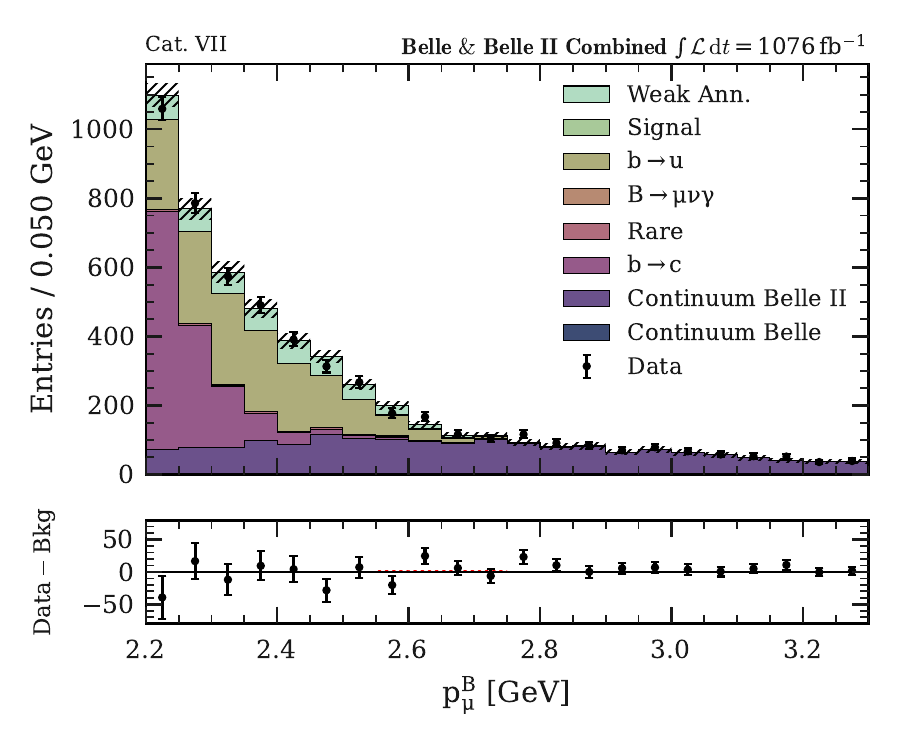}
    \includegraphics[height=0.23\textheight]{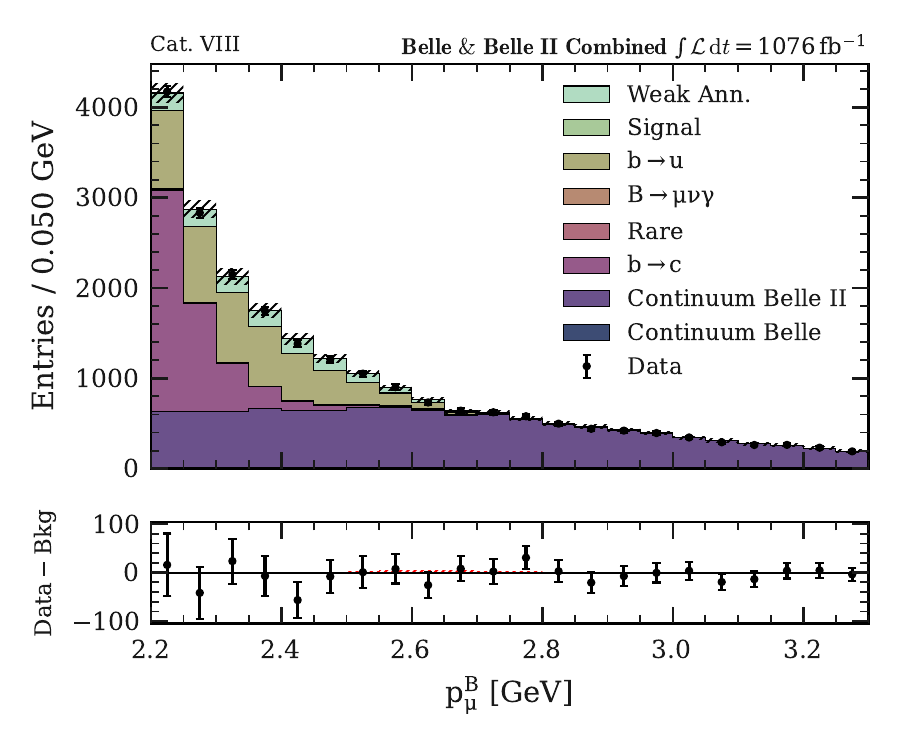}
    \caption{Distribution of the fitted muon momentum in the \(B\) rest frame for the eight categories, with an additional template for weak annihilation events included.}
    \label{fig:wa_fit_eight_cats}
\end{figure}

The post-fit distributions of the nominal and WA fits are shown in Figs.~\ref{fig:fit_eight_cats} and \ref{fig:wa_fit_eight_cats}. The residuals, defined as the data yield minus the prediction from all templates excluding $B^+ \to \mu^+ \nu_\mu$, include statistical uncertainties of the data only.

\begin{table}[htbp]
	\caption{The uncertainties on the measured \(\Delta\mathcal{B}(B \to X_u \ell^+ \nu_\ell)\) partial branching fraction are shown. The multiplicative systematic uncertainties arising from tracking, particle identification, \(B\bar{B}\) counting, control-channel efficiencies, and efficiency variations due to systematic effects are combined. The dominant contribution to the multiplicative systematic uncertainty comes from the control-channel efficiencies.}
\begin{tabular}{l|c}
    \hline \hline
	Source                                & Fractional \\ 
	     & uncertainty \\ \hline \hline 
    \textbf{Additive uncertainties}&\\ \hline
        $\quad  $$b \to u$ modeling&8.5\%\\ 
	$\quad \quad $$B \to \pi \ell^+ \nu_\ell$ FF&1.1\%\\ 
	$\quad \quad $$B \to \rho \ell^+ \nu_\ell$ FF&5.9\%\\ 
	$\quad \quad $$B^+ \to \omega \ell^+ \nu_\ell$ FF&2.8\%\\ 
	$\quad \quad $$B^+ \to \eta \ell^+ \nu_\ell$ FF&0.1\%\\ 
	$\quad \quad $$B^+ \to \eta' \ell^+ \nu_\ell$ FF&0.6\%\\ 
	$\quad \quad $$B \to \pi \ell^+ \nu_\ell$ BF&0.6\%\\ 
	$\quad \quad $$B \to \rho \ell^+ \nu_\ell$ BF&0.2\%\\ 
	$\quad \quad $$B^+ \to \omega \ell^+ \nu_\ell$ BF&0.2\%\\ 
	$\quad \quad $$B^+ \to \eta \ell^+ \nu_\ell$ BF&$<$ 0.1\%\\ 
	$\quad \quad $$B^+ \to \eta' \ell^+ \nu_\ell$ BF&$<$ 0.1\%\\ 
	$\quad \quad $$B \to X_u \ell^+ \nu_\ell$ BF&1.6\%\\ 
	$\quad \quad $DFN parameters&3.4\%\\ 
	$\quad \quad $Hybrid model&3.7\%\\ 
	$\quad \quad $MC sample size&0.3\%\\ 
    
	$\quad $Continuum modeling&0.5\%\\ 
	$\quad \quad$Shape correction&0.2\%\\ 
	$\quad \quad$MC sample size&0.5\%\\  
    
	$\quad $Rare decay modeling&$<$ 0.1\%\\ 
    
	$\quad $$B^+ \to \mu^+ \nu_\mu \gamma$ modeling&0.4\%\\     

	$\quad $$b \to c$ modeling&0.3\%\\ 
    
	$\quad B^+ \to \mu^+ \nu_\mu$ modeling&$<$ 0.1\%\\  \hline 
    \textbf{Multiplicative uncertainties}& 6.1\%\\  \hline  \hline 
	\textbf{Total systematic uncertainty}&10.8\%\\ 
	\textbf{Total statistical uncertainty}&1.7\%\\ 
    \hline \hline
\end{tabular}
    \label{table:sys_table_comb_Xu}
\end{table}

\twocolumngrid

\end{document}